\documentclass[10pt,journal,final]{IEEEtran}
\IEEEoverridecommandlockouts

\usepackage{amsfonts,amssymb}
\usepackage{ifpdf}
\usepackage{cite}
\usepackage{stfloats}
\usepackage{bm}
\usepackage{color,xcolor}
\usepackage{subfigure}

\ifCLASSINFOpdf
\usepackage[pdftex]{graphicx}
\graphicspath{{../pdf/}{../jpeg/}}
\DeclareGraphicsExtensions{.pdf,.jpeg,.png}
\else
\usepackage[dvips]{graphicx}

\graphicspath{{../eps/}}

\DeclareGraphicsExtensions{.eps}
\fi

\usepackage[cmex10]{amsmath}

\usepackage{algorithm}
\usepackage{algorithmic}
\ifCLASSOPTIONcompsoc
\else
\usepackage[caption=false,font=footnotesize]{subfig}
\fi

\usepackage{makecell}

\begin{document}	
\title{Integrated Sensing and Communication: Joint Pilot and Transmission Design}
\author{Meng~Hua,~\IEEEmembership{Member,~IEEE,}
	Qingqing~Wu,~\IEEEmembership{Senior Member,~IEEE,}
 Wen~Chen,~\IEEEmembership{Senior Member,~IEEE,}
 Abbas Jamalipour,~\IEEEmembership{Fellow,~IEEE,}
 Celimuge~Wu,~\IEEEmembership{Senior Member,~IEEE,}
and~Octavia A. Dobre,~\IEEEmembership{Fellow,~IEEE}
%
\thanks{M. Hua  is with the Department of Electronic Engineering, Shanghai Jiao Tong University,  Minhang 200240, China, and also  with the State Key Laboratory of Internet of Things for Smart City, University of Macau, Macao 999078, China (email: menghua@um.edu.mo).}
\thanks{	 Q. Wu and W. Chen are with the Department of Electronic Engineering, Shanghai Jiao Tong University, Minhang 200240, China (e-mail: qingqingwu@sjtu.edu.cn; wenchen@sjtu.edu.cn).}
\thanks{ A. Jamalipour is with the School of Electrical and Information Engineering, The University of Sydney, Sydney, NSW 2006, Australia (e-mail:
	a.jamalipour@ieee.org).}
\thanks{C. Wu is with the Department of Computer and
Network Engineering, The University of Electro-Communications, Tokyo
182-8585, Japan (e-mail: celimuge@uec.ac.jp).}
\thanks{O. A. Dobre is with the Faculty of Engineering and Applied
	Science, Memorial University, St. John’s, NL A1B 3X5, Canada  (email: odobre@mun.ca).}
}
\maketitle
\vspace{-1.2cm}
\begin{abstract}	
This paper studies a communication-centric integrated sensing and communication (ISAC)  system, where a multi-antenna base station (BS)  simultaneously performs downlink  
communication and  target detection.
A novel target detection and information transmission protocol is proposed, where the BS executes the channel
estimation and  beamforming successively and meanwhile jointly exploits the pilot sequences in the channel estimation stage and user  information in the transmission stage to assist target detection. We investigate the joint design of the pilot matrix, training duration, and transmit beamforming to maximize the probability of target detection, subject to the minimum  achievable rate required by the user. However, designing the optimal pilot matrix is rather challenging  since there is no closed-form expression of the detection probability with respect to the pilot matrix. To tackle this difficulty, we resort to designing the pilot matrix based on the information-theoretic criterion to maximize  the   mutual information (MI) between the received observations and BS-target channel coefficients  for target detection. We first derive the optimal pilot matrix for both channel estimation and target detection, and then  propose a unified pilot matrix structure  to balance minimizing the channel estimation error (MSE) and  maximizing  MI. Based on the proposed structure, a low-complexity successive refinement algorithm is proposed. In addition, we rigorously analyze  the impact of pilot length and pilot matrix on two fundamental  tradeoffs,  namely MSE-MI and Rate-MI.
Simulation results demonstrate that the proposed pilot matrix structure can well balance  the MSE-MI and the  Rate-MI tradeoffs, and show the significant region improvement of our   proposed  design as compared to other benchmark schemes. Furthermore, it is unveiled  that as the  communication channel is  more spatially correlated, 
the Rate-MI region can be further  enlarged.	
\end{abstract}

\section{Introduction}

Future emerging applications such as Internet of Things (IoT) smart cities will pose new requirements on future wireless communication networks \cite{zhang2022enabling, AHELEROFF2020101043,Chen2022irs}, which  not only require high-data-rate and low-latency communication services but also  additional high precision and high-resolution sensing functions. As shown by Statista organization, the number of IoT   devices   is predicated to grow from $13.1$ billion in 2022 to more than $29.4$ billion in 2030 \cite{onlineweb}.
To support such massive amount of smart IoT devices, the integrated sensing and communication (ISAC) technology is recently proposed, where the base station (BS) integrates the sensing and communication functions into a common platform and can operate two functions in the same frequency band \cite{Liu2022survey,Ma2020joint,luong2021radio,liu2020joint,chen2022isac,9997576}.

Different from the coexistence system where  the radar and communication hardware modules  are physically separated, they  are physically integrated  into  the ISAC system. Thus, several appealing advantages are introduced as follows \cite{cui2021integrating}. \textit{1) Ease of integration}: The majority of transmitter/receiver modules can be shared by radar  and communication systems, which makes it easy to implement from hardware perspectives; \textit{2) Integration gain}: The
components or resources in the ISAC system can be coupled to achieve more efficient
resource utilization such that the signal overhead can be reduced and the  spectral and  energy efficiency can be improved;
\textit{3) Coordination gain}: One can flexibly balance the  dual-functional performance via mutual assistance  such as jointly designing the waveforms. As such, a large number of works have paid attention to it in the literature, which can be classified into three paradigm directions of research, namely radar-centric design \cite{Hassanien2016dual,hassanien2016phase,wang2018sparse}, communication-centric design \cite{Sturm2011waveform,sit2011doppler,braun2014ofdm}, and joint design and optimization \cite{hua2022joint,joint2018liu,liu2018towards,hua2021optimal,luo2022joint}, on the ISAC system based on the different integration approaches. For example, in \cite{Hassanien2016dual}, the authors considered the radar system as the primary function and embedded communication signals into the radar waveform by controlling its amplitude and phase of the radar spatial side-lobe to convey information. In \cite{Sturm2011waveform}, the authors considered the existing communication transmitter hardware and   applied the orthogonal frequency-division multiplexing (OFDM) communication signals to realize the radar sensing functionality. The authors in  \cite{joint2018liu} proposed a new hardware architecture which is able  to  jointly design  communication and radar waveforms  to realize both communication and sensing functionalities. 

However, it is worth pointing out that none of the above works focused on the ISAC system design considering both the
downlink training and information transmission phases. It still remains unknown how   the pilot matrix,  the training duration, and the transmit beamformer  impact  communication and sensing. First,  since the second-order statistic of the channel state information (CSI) for the communication user channel and the target channel is in general different, the optimal pilot matrix for channel estimation may not 
be optimal for target sensing, and vice versa. Second,  if  a longer
training duration   is  used for improving the accuracy of  channel estimation, less time is left for data transmission, which gives rise to  a fundamental  tradeoff between channel estimation accuracy  and   data transmission. Furthermore, the  time allocated for channel estimation and information transmission will also impact the target sensing since the pilot signals and the information  signals are not the same. To be specific, the pilot matrix is deterministic and  remains unchanged over a whole channel coherence while the information signals are random and  vary  over different time slots. Third, the optimal transmit beamformer for  information transmission and target sensing is different since the narrow beam is expected   to focus all the energy  on   the communication user, while  a flat beam  is  desired for target sensing since the target location is unknown.     Therefore,  a unified resource allocation, i.e.,   space-time code/pilot matrix, training duration, and the transmit beamformer,  on the ISAC system should be thoroughly studied, which thus motivates  this work.
 We note that 
 \cite{maio2007design} studied  the optimal  space-time code design for radar detection, whereas  only the radar system was considered and its impact on the communication system still remains unknown. In addition, the authors in  \cite{8663356} unveiled the impact of  pilot training duration  on the communication system performance while  its impact on the radar system was not studied. However, their  insights will not be true in the ISAC system and  their proposed transceiver designs are also  no longer applicable  due to the   joint resource allocation.

\begin{figure}[!t]
	\centerline{\includegraphics[width=3.5in]{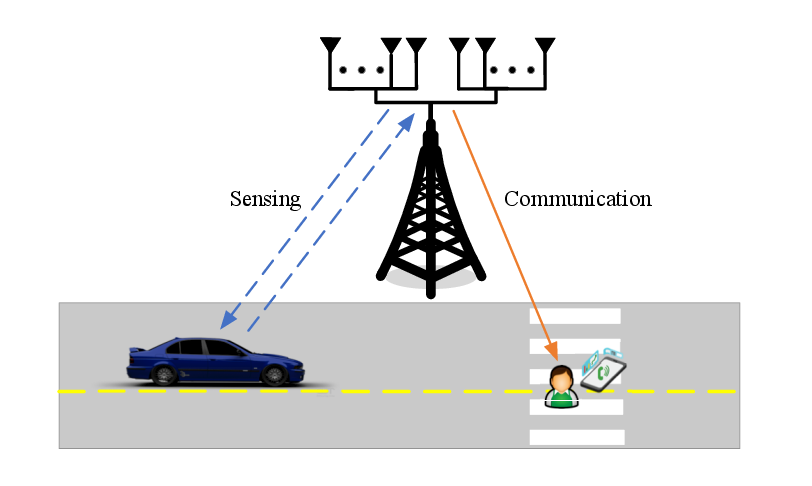}}
	\caption{A communication-centric ISAC system.} \label{fig1}
		\vspace{-0.5cm}
\end{figure}

As shown in Fig.~\ref{fig1}, we consider  a  communication-centric  frequency-division duplexing (FDD) ISAC  system with one BS, one potential target, and one communication user, where  the  BS  simultaneously performs downlink   communication and  target detection. This paper attempts to study the performance tradeoff between  communication and sensing based on the communication-centric ISAC systems where the BS leverages the communication waveforms to perform sensing. We study the resource allocation, namely  pilot training duration, transmit beamformer, and  space-time code/pilot matrix,  on the ISAC system to maximize the target detection probability while guaranteeing the communication system performance. Particularly, we unveil two tradeoffs, namely channel estimation error (MSE)-mutual information (MI) and Rate-MI, on the ISAC. It is worth mentioning that our work differs significantly from \cite{joint2022huang} in  four  aspects. First, \cite{joint2022huang}  studies 
 a time-division duplexing (TDD)  ISAC system where the BS sends downlink pilot signals for target detection and the user sends uplink pilot signals to the BS for channel estimation, while our work studies an FDD  ISAC system  where the BS sends downlink pilot signals  for target detection and the user performs channel estimation. Second, the objective of \cite{joint2022huang}  is to  enhance sensing  performance by exploiting joint burst sparsity and pilot design, while our work is to study the performance tradeoff  between communication and sensing under the
 pilot matrix, the training duration, and the transmit beamformer. Third, the proposed algorithm in \cite{joint2022huang} is not applicable to our considered optimization problem,  we propose an efficient algorithm to cater to the formulated optimization problem. Fourth,  \cite{joint2022huang} does not consider the communication performance, while our work considers  the impact of pilot duration and pilot signals on the ISAC system.
The contributions of his paper are summarized as follows:
\begin{itemize}
	\item  We propose a novel target detection and information transmission protocol, where both the channel estimation stage and information transmission stage are used to target detection. 
	 The closed-form formulas for the false alarm and the detection
	 probability  based on the generalized likelihood  ratio test (GLRT) are derived.
	Then,  a target  detection probability maximization optimization problem  is formulated by jointly optimizing pilot training duration, transmit beamformer, and  space-time code/pilot matrix,  subject to the minimum achievable rate required by the communication user. 	
	\item  To minimize the  channel estimation error, we find that
	the optimal pilot matrix is nonunitary with different power allocation. The analysis shows that  with a fixed power budget, a larger training duration leads to a smaller MSE, while the MSE will not be further reduced as the training duration is larger than the rank of the channel covariance matrix. In contrast, to maximize the target detection probability  based on the MI merit, we find that   the 
	optimal pilot matrix is unitary with equal power allocation.  We prove that with a fixed power budget, a larger training duration leads to a larger MI, while the MI will not be further increased as the training duration is larger than the number of transmit antennas. 
	Based on these observations, a novel pilot matrix structure is proposed that balances the tradeoff between the MSE  and the MI. Then, we solve the resulting optimization problem based on the block coordinate descent (BCD) approach, where the beamformer and pilot matrix are alternately optimized.
	\item Simulation results show that the proposed  design is  capable of
	substantially improving the target detection probability, Rate-MI region, and MSE-MI region  compared to the benchmark schemes. In addition, 
	  several interesting insights are unveiled. First,  the 
	proposed pilot structure can well balance  the  communication  performance  and the  sensing performance, and show the superiority of the proposed pilot structure over the discrete Fourier transform (DFT) matrix and Gaussian-based matrix. Second,  the optimal training duration for maximizing the Rate-MI region is not equal to the rank of the channel covariance matrix. Third, 	the Rate-MI region can be further enlarged for a more spatially correlated communication channel.
\end{itemize} 
The rest of this paper is organized as follows. Section II introduces the system model and the problem formulation  for the considered communication-centric ISAC. In Section III, the novel pilot matrix structure is proposed and a BCD-based  algorithm  is further introduced to solve the resulting optimization problem.
Numerical results are provided in Section IV and the paper is concluded in Section V.

\emph{Notations}: Boldface upper-case and lower-case  letters denote matrices and   vectors, respectively. ${{\bf{1}}_{L}}$ represents a vector of  all ones  with the length of $L$.  ${\mathbb C}^ {d_1\times d_2}$ stands for the set of  complex $d_1\times d_2$  matrices. For a complex-valued vector $\bf x$, ${\left\| {\bf x} \right\|}$ represents the  Euclidean norm of $\bf x$, and ${\rm diag}(\bf x) $ denotes a diagonal matrix whose main diagonal elements are extracted from vector $\bf x$.
${\left(  \cdot  \right)^T},{\left(  \cdot  \right)^*}$, ${\left(  \cdot  \right)^H}$, and  ${\left(  \cdot  \right)^{\dag}}$stand for  the transpose operator,  conjugate operator,  conjugate transpose, and pseudo inverse operator, respectively.  ${\left\| {\bf{X}} \right\|_F}$  represents the  Frobenius norm of ${\bf{X}}$, ${\bf{X}} \succeq {\bf{0}}$ indicates that matrix $\bf X$ is positive semi-definite, and ${\bf{X}}\left( {1:L} \right)$ stands for the matrix containing the first $L$ columns of ${\bf{X}}$.
A circularly symmetric complex Gaussian (CSCG) vector $\bf x$ with mean $ \bm {\mu}$ and  covariance matrix ${\sigma ^2}{\bf{I}}$ is denoted by ${\bf x} \sim {\cal CN}\left( {{{\bm \mu }},{{\sigma^2{\bf{I}} }}} \right)$. ${\mathbb Z}^+$ represents the positive integer notation. $ \otimes $ denotes the Kronecker product operator and  ${\cal O}\left(  \cdot  \right)$ is the big-O computational complexity notation.

\section{System Model and Problem Formulation}
Consider a narrow-band  ISAC system consisting of one BS, one single-antenna user, and one potential target, as shown in Fig.~\ref{fig1}.\footnote{Although we consider one communication user, our proposed algorithm can be readily applicable to the case with multiple users since the channel estimation is performed on the user side and there is no difference in the channel estimation between the single user and multiple users. In addition, although the iterative  GLRT can be applied for multi-target detection \cite{4655353}, the impact of pilot design on the multi-target case is difficult to analyse, which requires non-trivial efforts and we would like to leave it as our future work.}  
The  BS is equipped with $N_t+N_r$ antennas, of which   $N_t$  transmit antennas are used for simultaneously  serving  communication users and sensing   radar targets in the same frequency band, while   $N_r$ receive  antennas  are dedicated to receiving the echo signals reflected by the  target. We consider a quasi-static flat-fading channel in which the CSI 
remains unchanged in a channel coherence frame, but may change in the subsequent frames.
 Note that  the frames of interest  have the same channel fading statistical distribution so that the same pilot training sequences and the training duration can be applied for all the frames. 
 Without loss of generality, we denote   the channel coherence duration of each frame by $T_c$ (in symbols).

\begin{figure}[!t]
	\centerline{\includegraphics[width=3.2in]{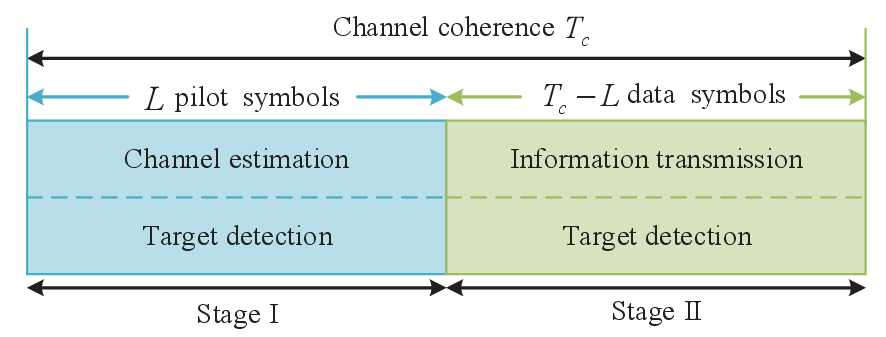}}
	\caption{Target detection  and information transmission protocol.} \label{fig2}
		\vspace{-0.5cm}
\end{figure}
We consider an FDD system where  the BS sends the  downlink pilot sequences to the user for channel estimation, which then feeds back the CSI to  the BS.\footnote{Note that for the  time-division duplex (TDD) mode,  the BS  simultaneously receives the   pilot sequences transmitted by the user  and the  echo signals reflected by the target. The performance of both  channel estimation and target detection will be significantly deteriorated due to the  mutual interference. As a result, the TDD mode is not considered here.}   
The target detection  and information transmission protocol is shown in Fig.~\ref{fig2}, in which 
one channel coherence time  is divided into two stages, namely stage I and stage II.\footnote{We propose a novel and efficient protocol, where the BS can perform sensing in both the channel estimation stage and the information transmission stage. Thus,  the resources such as frequency and time  are fully exploited.} In  stage I, the BS transmits $L$   pilot symbols for the estimation and meanwhile these  pilot symbols are  used for target detection. In stage II, the BS transmits to the user $T_c-L$ data  symbols, which are also used for target detection. Under this protocol, the target detection is fully utilized  during one channel coherence time.
For convenience, we denote the sets of  BS transmit antennas, BS receive antennas,  pilot symbols, and channel coherence interval   as ${\cal N}_t$,  ${\cal N}_r$,  $\cal L$, and ${\cal T}_c$, respectively.
In this paper, our analysis is based on the spatially correlated Rician channel model for the communication channel and the uncorrelated channel model for the radar channel. The analysis of the other channels such as the multipath channel model is interesting but requires non-trivial efforts and we would like to leave them as our future work. 
\subsection{Communication-Centric Target Detection}
In stage I, the signal received at the $i$th  BS receive antenna during the $l$th symbol is given by
\begin{align}
{y_{r,i}}\left[ l \right] =\sum\limits_{j = 1}^{{N_t}} {{g_{i,j}}{x_{c,j}}\left[ l \right]}  + {n_{r,i}}\left[ l \right], i\in {\cal N}_r,l \in {\cal L},
\end{align}
where  ${x_{c,j}}\left[ l \right]$ denotes the $l$th pilot symbol transmitted by the $j$th BS transmit antenna,  $g_{i,j}$ stands for the round-trip channel coefficient  between the $j$th BS transmit antenna
and the $i$th BS receive antenna,\footnote{$g_{i,j}$ accounts for both the  channel propagation and the radar-cross section of the target.  The target is in general composed of an infinite number
		of random, isotropic and independent scatterers, we model the  radar-cross section as the Gaussian random variable caused by this  fluctuation \cite{1597550}.} and ${n_{r,i}}\left[ l \right] \sim {\cal CN}\left( {0,\sigma _r^2} \right)$ stands for the additive white Gaussian noise. Note that in the sequel, we assume that  round-trip channel coefficients $g_{i,j}$'s are independent by assuming that the   transmit antennas and  receive antennas at the BS are sufficiently spaced so that the angle diversity can be    explored.  In addition, we consider a  Swerling-I target model, where the channel coefficient
follows identically distributed CSCG with mean $0$ and variance ${\delta _g^2}$, i.e., ${g_{i,j}} \sim {\cal CN}\left( {0,\delta _g^2} \right)$ \cite{maio2007design}.

Upon collecting $L$ symbols at the $i$th BS receive antenna and defining ${{\bf{y}}_{r,i}^{\rm I}} = {\left[ {{y_{r,i}}\left[ 1 \right], \ldots ,{y_{r,i}}\left[ L \right]} \right]^T}$, we can rewrite it as a vector form given by 
\begin{align}
{{\bf{y}}_{r,i}^{\rm I}} = {{\bf{X}}_c}{{\bf{g}}_i} + {{\bf{n}}_{r,i}^{\rm I}}, i\in {\cal N}_r,
\end{align}
where ${{\bf{X}}_c} = \left[ {\begin{array}{*{20}{c}}
	{{x_{c,1}}\left[ 1 \right]}& \cdots &{{x_{c,{N_t}}}\left[ 1 \right]}\\
	\vdots & \ddots & \vdots \\
	{{x_{c,1}}\left[ L \right]}& \cdots &{{x_{c,{N_t}}}\left[ L \right]}
	\end{array}} \right]$, ${{\bf{g}}_i} = {\left[ {{g_{i,1}}, \ldots ,{g_{i,{N_t}}}} \right]^T}$, and ${{\bf{n}}_{r,i}^{\rm I}} = {\left[ {{n_{r,i}}\left[ 1 \right], \ldots ,{n_{r,i}}\left[ L \right]} \right]^T}$. 

Similar to  stage I, the pilot symbols  in  stage II are replaced by the  data symbols for transmission. As such, we can write the signal received at the $i$th BS receive antenna after collecting $T_c-L$ data symbols in stage II as 
\begin{align}	
{\bf{y}}_{r,i}^{{\rm{II}}} &=  {\rm{diag}}\left( {{x_t}\left[ {L{\rm{ + }}1} \right], \ldots ,{x_t}\left[ {{T_c}} \right]} \right){{\bf{1}}_{{T_c} - L}}{{\bf{w}}^T}{{\bf{g}}_i} + {\bf{n}}_{r,i}^{{\rm{II}}}\notag\\
& =  {{\bf{X}}_t}{{\bf{g}}_i} + {\bf{n}}_{r,i}^{{\rm{II}}}, i\in {\cal N}_r,
\end{align}
where  ${{\bf{X}}_t} = {\rm{diag}}\left( {{x_t}\left[ {L{\rm{ + }}1} \right], \ldots ,{x_t}\left[ {{T_c}} \right]} \right){{\bf{1}}_{{T_c} - L}}{\bf{w}}^T$,  ${{x_t}\left[ j \right]}$ denotes the $j$th data symbol, ${\bf{w}} \in {{\mathbb C}^{{N_t} \times 1}}$ stands for the transmit beamformer,
and ${\bf{n}}_{r,i}^{{\rm{II}}}$ denotes the received white Gaussian noise satisfying ${\bf{n}}_{r,i}^{{\rm{II}}} \sim {\cal CN}\left( {{{\bf{0}}_{\left( {{T_c} - L} \right) \times 1}},\sigma _r^2{{\bf{I}}_{{T_c} - L}}} \right)$.

Based on the presence (hypothesis ${{\cal H}_1}$) or absence (hypothesis ${{\cal H}_0}$) of the target, a binary hypothesis test over one channel coherence time  is formulated as 
\begin{align}
&{{\cal H}_0}:\;\;\;{\mkern 1mu} \left\{ {\begin{array}{*{20}{l}}
	{{\bf{y}}_{r,i}^{\rm{I}} = {\bf{n}}_{r,i}^{\rm{I}}, i\in {\cal N}_r, \;\;\;{\mkern 1mu} {\rm{stage~I}},}\\
	{{\bf{y}}_{r,i}^{{\rm{II}}} = {\bf{n}}_{r,i}^{{\rm{II}}},i\in {\cal N}_r, \;\;\;{\mkern 1mu} {\rm{stage~II}}.}
	\end{array}} \right. \notag\\
&{{\cal H}_1}:\;\;\;\left\{ {\begin{array}{*{20}{l}}
	{{\bf{y}}_{r,i}^{\rm{I}} = {{\bf{X}}_c}{{\bf{g}}_i} + {\bf{n}}_{r,i}^{\rm{I}},i\in {\cal N}_r,\;\;\;{\mkern 1mu} {\rm{stage~I}},}\\
	{{\bf{y}}_{r,i}^{{\rm{II}}} =  {\bf{X}}_t{{\bf{g}}_i} + {\bf{n}}_{r,i}^{{\rm{II}}},i\in {\cal N}_r,\;\;\;{\mkern 1mu} {\rm{stage ~II}}.}
	\end{array}} \right. \label{hypothesis_1}
\end{align}
It can be readily checked that the hypothesis testing problem in \eqref{hypothesis_1} belongs to a model change detection problem, where  the mean value will jump with the change of time (see stage I and stage II in ${\cal H}_1$). Define ${{\bf{y}}_{r,i}} = {\left[ {{\bf{y}}_{r,i}^{{\rm{I,}}T}{\kern 1pt} {\kern 1pt} {\kern 1pt} {\kern 1pt} {\kern 1pt} {\kern 1pt} {\bf{y}}_{r,i}^{{\rm{II,}}T}} \right]^T}$ and ${{\bf{n}}_{r,i}} = {\left[ {{\bf{n}}_{r,i}^{{\rm{I,}}T}{\kern 1pt} {\kern 1pt} {\kern 1pt} {\kern 1pt} {\kern 1pt} {\kern 1pt} {\bf{n}}_{r,i}^{{\rm{II,}}T}} \right]^T}$. Then, we can rewrite \eqref{hypothesis_1} in a more compact form given by 
\begin{align}
&{{\cal H}_0}:\;\;\;{\mkern 1mu} {{\bf{y}}_{r,i}} = {{\bf{n}}_{r,i}}, i\in {\cal N}_r, \notag\\
&{{\cal H}_1}:\;\;\;{{\bf{y}}_{r,i}} = {\bf{X}}{{\bf{g}}_i} + {{\bf{n}}_{r,i}}, i\in {\cal N}_r, \label{hypothesis_2}
\end{align}
where ${\bf{X}} = {\left[ {{\bf{X}}_c^T{\kern 1pt} {\kern 1pt} {\kern 1pt} {\bf{X}}_t^T} \right]^T}$.

Since the prior information  about ${\bf g}_i$'s is unknown, the traditional Neyman–Pearson criterion cannot be applied. Instead, the generalized likelihood
ratio test (GLRT) is adopted,  which   replaces the unknown
parameters with their maximum likelihood (ML)
estimates under each hypothesis. Specifically, under the hypothesis test in   \eqref{hypothesis_2}, the GLRT decides ${\cal H}_1$ or ${\cal H}_0$ as follows: 
\begin{align}
\frac{{\mathop {\max }\limits_{\left\{ {{{\bf{g}}_i}} \right\}} f\left( {{{\bf{y}}_{r,1}} \ldots ,{{\bf{y}}_{r,{N_r}}};{{\bf{g}}_1}, \ldots ,{{\bf{g}}_{{N_r}}}|{{\cal H}_1}} \right)}}{{f\left( {{{\bf{y}}_{r,1}} \ldots ,{{\bf{y}}_{r,{N_r}}}|{{\cal H}_0}} \right)}}  \mathop{\gtrless}_{{\cal H}_0}^{{\cal H}_1} {\Gamma _{{{\rm th}}}}, \label{GLRT}
\end{align}
where ${\Gamma _{{{\rm th}}}}$ denotes the decision threshold,  $f\left( {{{\bf{y}}_{r,1}} \ldots ,{{\bf{y}}_{r,{N_r}}};{{\bf{g}}_1}, \ldots ,{{\bf{g}}_{{N_r}}}|{{\cal H}_1}} \right)$ and $f\left( {{{\bf{y}}_{r,1}} \ldots ,{{\bf{y}}_{r,{N_r}}}|{{\cal H}_0}} \right)$ denote the probability density functions  (PDFs) of the data under hypotheses ${\cal H}_1$  and ${\cal H}_0$ from $N_r$ receive antennas, which are respectively given by 
\begin{align}
&f\left( {{{\bf{y}}_{r,1}} \ldots ,{{\bf{y}}_{r,{N_r}}}|{{\cal H}_0}} \right) = \notag\\
& \qquad \qquad \frac{1}{{{\pi ^{{T_c}{N_r}}}\sigma _r^{2{T_c}{N_r}}}}\exp \left( { - \frac{1}{{\sigma _r^2}}\sum\limits_{i = 1}^{{N_r}} {{{\left\| {{{\bf{y}}_{r,i}}} \right\|}^2}} } \right), \label{pdfh0}
\end{align}
\begin{align}
&f\left( {{{\bf{y}}_{r,1}} \ldots ,{{\bf{y}}_{r,{N_r}}};{{\bf{g}}_1}, \ldots ,{{\bf{g}}_{{N_r}}}|{{\cal H}_1}} \right) = \notag\\
&\qquad \frac{1}{{{\pi ^{{T_c}{N_r}}}\sigma _r^{2{T_c}{N_r}}}}\exp \left( { - \frac{1}{{\sigma _r^2}}\sum\limits_{i = 1}^{{N_r}} {{{\left\| {{{\bf{y}}_{r,i}} -{\bf{X}}{{\bf{g}}_i}} \right\|}^2}} } \right). \label{pdfh1}
\end{align}
Substituting \eqref{pdfh0} and \eqref{pdfh1} into \eqref{GLRT},   we can further simplify \eqref{GLRT} to
\begin{align}
\frac{1}{{\sigma _r^2}}\left(\sum\limits_{i = 1}^{{N_r}} {{{\left\| {{{\bf{y}}_{r,i}}} \right\|}^2}}  - \mathop {\min }\limits_{\left\{ {{{\bf{g}}_i}} \right\}} \sum\limits_{i = 1}^{{N_r}} {{{\left\| {{{\bf{y}}_{r,i}} - {\bf{X}}{{\bf{g}}_i}} \right\|}^2}}\right)  \mathop{\gtrless}_{{\cal H}_0}^{{\cal H}_1} {{\hat \Gamma }_{{{\rm th}}}},\label{GLRT_v1}
\end{align}
where ${{\hat \Gamma }_{{{ \rm th}}}} = \ln {\Gamma _{{{\rm th}}}}$.

Let ${{{\bf{\hat g}}}_i}$ be the  ML estimate of  ${{{\bf{ g}}}_i}$ under hypotheses ${\cal H}_1$, $i\in {\cal N}_r$.  
%
Taking the first-order derivative of ${{{\left\| {{{\bf{y}}_{r,i}} - {\bf{X}}{{\bf{g}}_i}} \right\|}^2}}$ with respect to (w.r.t.) ${{{\bf{g}}_i}}$ and setting it to zero, we have 
\begin{align}
{{{\bf{\hat g}}}_i} = {\left( {{{\bf{X}}^H}{\bf{X}}} \right)^{ \dag}}{{\bf{X}}^H}{{\bf{y}}_{r,i}},i\in {\cal N}_r. \label{targetchanneloptiaml}
\end{align}
Then, substituting \eqref{targetchanneloptiaml} into \eqref{GLRT_v1},  we have 
\begin{align}
\frac{1}{{\sigma _r^2}}\left(\sum\limits_{i = 1}^{{N_r}} {{\bf{y}}_{r,i}^H} {\bf{X}}{\left( {{{\bf{X}}^H}{\bf{X}}} \right)^{ \dag }}{{\bf{X}}^H}{{\bf{y}}_{r,i}}\right)\mathop{\gtrless}_{{\cal H}_0}^{{\cal H}_1}{{\hat \Gamma }_{{{\rm th}}}}.\label{GLRT_v2}
\end{align}
In this paper, we assume that the pilot-based matrix ${\bf{X}}_c$ is a full-rank matrix to achieve the maximum spatial multiplexing gain. In addition, based on the fact that the pilot symbols and data symbols are independent in general, we have the following lemma.

\textbf{\emph{Lemma 1:}} If ${\bf{X}}_c$ is a full-rank matrix and ${{\bf{X}}_t} \ne {\bf{0}}$, i.e., ${\rm{rank}}\left( {{{\bf{X}}_c}} \right) = \min \left( {L,{N_t}} \right)$, the rank of $\bf X$ is given by   ${\rm{rank}}\left( {\bf{X}} \right) = \min \left( {L + 1,{N_t}} \right)$.

\hspace*{\parindent}\textit{Proof}:  This can be directly verified from the definition and is thus omitted here. \hfill\rule{2.7mm}{2.7mm}

\textbf{\emph{Lemma 2:}} ${\rm{rank}}\left( {{\bf{X}}{{\left( {{{\bf{X}}^H}{\bf{X}}} \right)}^\dag }{{\bf{X}}^H}} \right) = {\rm{rank}}\left( {\bf{X}} \right)$.

\hspace*{\parindent}\textit{Proof}:   Please refer to Appendix~\ref{appendix1}. \hfill\rule{2.7mm}{2.7mm}

\textbf{\emph{Lemma 3:}} ${{\bf{X}}{{\left( {{{\bf{X}}^H}{\bf{X}}} \right)}^\dag }{{\bf{X}}^H}}$ is idempotent and   has  $L+1$ eigenvalues  of one and $T_c-L -1$  eigenvalues of zero  when  $L\le N_t-1$, while it has   $N_t$ eigenvalues  of one and $T_c-N_t$  eigenvalues of zero  when  $L> N_t-1$.

\hspace*{\parindent}\textit{Proof}:   Please refer to Appendix~\ref{appendix2}. \hfill\rule{2.7mm}{2.7mm}
\subsubsection{Probability of False Alarm}
Under hypothesis ${\cal H}_0$, the received signal at the $i$th receiver given in \eqref{hypothesis_2}  satisfies  ${{\bf{y}}_{r,i}} \sim {\cal CN}\left( {{\bf{0}},\sigma _r^2{{\bf{I}}_{{T_c}}}} \right)$. 
Let ${{\bf{z}}_{r,i}} = {\bf{U}}^H{{\bf{y}}_{r,i}}$ and recall that ${\bf{X}}{\left( {{{\bf{X}}^H}{\bf{X}}} \right)^\dag }{{\bf{X}}^H} = {{\bf{U}}_{\bf{x}}}{\bf{U}}_{\bf{x}}^H$ (see in Appendix B), it can readily follow that ${{\bf{z}}_{r,i}} \sim {\cal CN}\left( {0,\sigma _r^2{{\bf{I}}_{{\min \left( {L + 1,{N_t}} \right)}}}} \right)$ based on the fact that an orthogonal transformation will not change the distribution of  ${{\bf{y}}_{r,i}}$. Then, the left-hand side of \eqref{GLRT_v2} can be rewritten as 
\begin{align}
\frac{1}{{\sigma _r^2}}\left( {\sum\limits_{i = 1}^{{N_r}} {{\bf{y}}_{r,i}^H} {\bf{X}}{{\left( {{{\bf{X}}^H}{\bf{X}}} \right)}^\dag }{{\bf{X}}^H}{{\bf{y}}_{r,i}}} \right) = \sum\limits_{i = 1}^{{N_r}} {{{\left( {\frac{{{{\bf{z}}_{r,i}}}}{{{\sigma _r}}}} \right)}^H}\frac{{{{\bf{z}}_{r,i}}}}{{{\sigma _r}}}} ,  \label{GLRT_v3}
\end{align}
with ${{\bf{z}}_{r,i}}/{\sigma _r} \sim {\cal CN}\left( {{\bf 0},{{\bf{I}}_{{\min \left( {L + 1,{N_t}} \right)}}}} \right)$. It can be readily verified that  \eqref{GLRT_v3}   follows  the central chi-squared distribution since 
it has  a sum of  the  squares of $2{N_r}\min \left( {L + 1,{N_t}} \right)$ independently real Gaussian random variables, each of which satisfies  zero mean and variance $1/2$. 
As such, the PDF of \eqref{GLRT_v3}  is given by \cite{proakis1998digital}
\begin{align}
{p_Z}\left( {z|{{\cal H}_0}} \right) = \frac{{{z^{\frac{1}{2}\min \left( {L + 1,{N_t}} \right) - 1}}}}{{\int_0^{ + \infty } {{t^{\frac{1}{2}\min \left( {L + 1,{N_t}} \right) - 1}}} {e^{ - t}}dt}}{e^{ - z}},z \ge 0.
\end{align}

Then, the probability of false alarm can be derived as \cite{proakis1998digital}
\begin{align}
{P_{{\rm{fa}}}} &= \int_{{{\hat \Gamma }_{{\rm{th}}}}}^{ + \infty } {{p_Z}\left( {z|{{\cal H}_0}} \right)dz} \notag\\
& = {e^{ - {{\hat \Gamma }_{{\rm{th}}}}}}\sum\limits_{k = 0}^{{N_r}\min \left( {L + 1,{N_t}} \right) - 1} {\frac{1}{{k!}}} \hat \Gamma _{{\rm{th}}}^k.
\end{align}
\subsubsection{Probability of Detection}
Under hypothesis ${\cal H}_1$,  the received signal at the $i$th receiver given in \eqref{hypothesis_2}  satisfies 
${{\bf{y}}_{r,i}}\sim {\cal CN}\left( {{\bf{X}}{{\bf{g}}_i},\sigma _r^2{{\bf{I}}_{{T_c}}}} \right)$. Similar to the hypothesis ${\cal H}_0$ case, define ${{\bf{z}}_{r,i}} = {\bf{U}}^H{{\bf{y}}_{r,i}}$, it follows that ${{\bf{z}}_{r,i}}\sim {\cal CN}\left( {{\bf{U}}_{\bf{x}}^H{\bf{X}}{{\bf{g}}_i},\sigma _r^2{{\bf{I}}_{\min \left( {L + 1,{N_t}} \right)}}} \right)$.  This indicates that each entry of ${{\bf{z}}_{r,i}}$ follows the Gaussian distribution and has the same variance but with  different means. 
As such, $\frac{1}{{\sigma _r^2}}\left( {\sum\limits_{i = 1}^{{N_r}} {{\bf{y}}_{r,i}^H} {\bf{X}}{{\left( {{{\bf{X}}^H}{\bf{X}}} \right)}^\dag }{{\bf{X}}^H}{{\bf{y}}_{r,i}}} \right)$ follows the non-central chi-squared distribution with the PDF given by  
\begin{align}
{p_Z}\left( {z|{{\cal H}_1}} \right) &= {\left( {\frac{z}{{{s^2}}}} \right)^{\left( {{N_r}\min \left( {L + 1,{N_t}} \right) - 1} \right)/2}}{e^{ - \left( {{s^2} + z} \right)}}\notag\\
&\times {I_{{N_r}\min \left( {L + 1,{N_t}} \right) - 1}}\left( {2\sqrt z s} \right), \label{GLRT_v4}
\end{align}
where  ${I_\alpha }\left( {2\sqrt z s} \right)$ represents the  Bessel function of the first kind of order $\alpha$ and the noncentrality parameter ${s^2}$  is given by 
\begin{align}
{s^2}& = \sum\limits_{i = 1}^{{N_r}} {{{\left\| {{\bf{U}}_{\bf{x}}^H{\bf{X}}{{\bf{g}}_i}} \right\|}^2}/\sigma _r^2} \notag\\
& \overset{(a)}{=} \sum\limits_{i = 1}^{{N_r}} {\left\| {{\bf{U}}_{\bf{x}}^H{{\bf{U}}_{\bf{x}}}{{\bf{\Sigma }}_{\bf{x}}}{\bf{V}}_{\bf{x}}^H{{\bf{g}}_i}} \right\|_F^2/\sigma _r^2}  \notag\\
&=  \sum\limits_{i = 1}^{{N_r}} {{\rm{tr}}\left( {{{\bf{\Sigma }}_{\bf{x}}}{\bf{V}}_{\bf{x}}^H{{\bf{g}}_i}{\bf{g}}_i^H{{\bf{V}}_{\bf{x}}}{{\bf{\Sigma }}_{\bf{x}}}} \right)/\sigma _r^2} \notag\\
&= \sum\limits_{i = 1}^{{N_r}} {{\rm{tr}}\left( {{\bf{X}}{{\bf{g}}_i}{\bf{g}}_i^H{{\bf{X}}^H}} \right)/\sigma _r^2}  =\left\| {\frac{1}{{{\sigma _r}}}{\bf{XG}}} \right\|_F^2,
\end{align}
where  equality (a) holds due to ${\bf{X}} = {{\bf{U}}_{\bf{x}}}{{\bf{\Sigma }}_{\bf{x}}}{\bf{V}}_{\bf{x}}^H$ defined in Appendix A.

As a consequence, the probability of detection can be derived as \cite{proakis1998digital} 
\begin{align}
{P_d} &= \int_{{{\hat \Gamma }_{{\rm{th}}}}}^{ + \infty } {{p_Z}\left( {z|{{\cal H}_1}} \right)dz} \notag\\
&={Q_{{N_r}\min \left( {L + 1,{N_t}} \right)}}\left( {\sqrt 2 s,\sqrt {2{{\hat \Gamma }_{{\rm{th}}}}} } \right)\notag\\
&={Q_{{N_r}\min \left( {L + 1,{N_t}} \right)}}\left( {\sqrt 2 {{\left\| {\frac{1}{{{\sigma _r}}}{\bf{XG}}} \right\|}_F},\sqrt {2{{\hat \Gamma }_{{\rm{th}}}}} } \right),
\end{align}
where ${Q_m}\left( {:,:} \right)$ is the generalized Marcum  $Q$ function of
order $m$ and ${\bf{G}} = \left( {{{\bf{g}}_1}, \ldots ,{{\bf{g}}_{{N_r}}}} \right)$.
\subsection{Channel Estimation and Information Transmission}
\subsubsection{Channel Estimation} Denote by  ${\bf{h}} \in {{\mathbb C}^{{N_t} \times 1}}$ the communication channel between the BS and the user.  To estimate $\bf h$, $L$ pilot sequences, i.e., ${\bf X}_c$ are applied. Then, the  signal received during stage I can be compactly written as
\begin{align}
{{\bf{y}}_p} =  {{\bf{X}}_c}{\bf{h}} + {{\bf{n}}_p}, \label{channel_est_1}
\end{align}
where ${{\bf{n}}_p} \sim {\cal CN}\left( {{\bf 0},\sigma _p^2{\bf I}_{L\times 1}} \right)$ represents the white Gaussian noise received by the user.

Based on the  signals received in \eqref{channel_est_1}, the minimum
mean square error (MMSE) estimator is applied for estimating $\bf h$. Specifically, the 
MMSE estimator  is given by 
\begin{align}
\!\!\!J\left( {\bf{F}} \right) = \mathop {\min }\limits_{\bf{F}} {\mathbb E}\left\{ {\left\| {{\bf{h}} - {\bf{\hat h}}} \right\|^2} \right\} =\mathop {\min }\limits_{\bf{F}} {\mathbb E}\left\{ {\left\| {{\bf{h}} - {\bf{F}}{{\bf{y}}_p}} \right\|^2} \right\}, \label{channel_est_2}
\end{align}
where ${\bf{\hat h}}$  denotes the estimated channel and  ${\bf{F}} \in {{\mathbb  C}^{{N_t} \times L}}$ is  a matrix to be optimized for minimizing the mean square    error (MSE) of the channel estimation.
By taking the first-order derivative of ${{{\left\| {{\bf{h}} - {\bf{F}}{{\bf{y}}_p}} \right\|}^2}}$ w.r.t. ${\bf{F}}$ and setting it to zero, the optimal ${\bf{F}} \in {{\mathbb  C}^{{N_t} \times L}}$ can be obtained as 
\begin{align}
{{\bf{F}}^{{\rm{opt}}}} = {{\bf{R}}_{\bf{h}}}{\bf{X}}_c^H{\left( {{{\bf{X}}_c}{{\bf{R}}_{\bf{h}}}{\bf{X}}_c^H + \sigma _p^2{{\bf{I}}_L}} \right)^{ - 1}},
\end{align}
where ${{\bf{R}}_{\bf{h}}} = {\mathbb E}\left\{ {{\bf{h}}{{\bf{h}}^H}} \right\}$.
Then, 
the estimation of $\bf h$ is given by 
\begin{align}
{\bf{\hat h}} = {{\bf{F}}^{{\rm{opt}}}}{{\bf{y}}_p}= {{\bf{R}}_{\bf{h}}}{\bf{X}}_c^H{\left( {{{\bf{X}}_c}{{\bf{R}}_{\bf{h}}}{\bf{X}}_c^H + \sigma _p^2{{\bf{I}}_L}} \right)^{ - 1}}{{\bf{y}}_p},  \label{channel_est_3}
\end{align}
with covariance matrix of ${\bf{\hat h}}$ given by 
\begin{align}
{{\bf{R}}_{{\bf{\hat h}}}}={{\bf{R}}_{\bf{h}}}{\bf{X}}_c^H\left( {{{\bf{X}}_c}{{\bf{R}}_{\bf{h}}}{\bf{X}}_c^H + \sigma _p^2{{\bf{I}}_L}} \right){{\bf{X}}_c}{{\bf{R}}_{\bf{h}}}.
\end{align}
Substituting \eqref{channel_est_3} into \eqref{channel_est_1}, the MSE  of  the MMSE estimator is given by
\begin{align}
\!\!{J_{{\mathop{\rm mmse}\nolimits} }}\left({{{\bf{X}}_c}} \right)& = {\rm{tr}}\left( {{{\bf{R}}_{\bf{h}}} - {{\bf{R}}_{\bf{h}}}{\bf{X}}_c^H{{\left( {{{\bf{X}}_c}{{\bf{R}}_{\bf{h}}}{\bf{X}}_c^H + \sigma _p^2{{\bf{I}}_L}} \right)}^{ - 1}}{{\bf{X}}_c}{{\bf{R}}_{\bf{h}}}} \right)\notag\\
&=\sigma _p^2{\rm{tr}}\left( {{{\bf{R}}_{\bf{h}}}{{\left( {{\bf{X}}_c^H{{\bf{X}}_c}{{\bf{R}}_{\bf{h}}} + \sigma _p^2{{\bf{I}}_{N_t}}} \right)}^{ - 1}}} \right).\label{J_mmse}
\end{align}

\subsubsection{Information Transmission}
In  stage II,  the signal detection procedure at the user 
is based on the estimated channel ${{\bf{\hat h}}}$.  The  signal received at the
user is rewritten as
\begin{align}
{y_t} &=  {{\bf{h}}^T}{\bf{w}}{x_t} + {{\bf{n}}_t}\notag\\
& =  {{{\bf{\hat h}}}^T}{\bf{w}}{x_t} + \underbrace {{\bf{h}}_e^T{\bf{w}}{x_t}}_{{\rm{channel~ estimation~ error}}} + ~{{\bf{n}}_t},
\end{align}
where  ${ {\bf{h}}_e^T{\bf{w}}{x_t}}$ is the additional interference caused by the channel
estimation error. Then, the average achievable rate in
bits/second/Hertz (bps/Hz) is  given by \cite{Samardzija2003pilot}\footnote{We focus on the beamformer design based on  the second-order statistic of CSI to reduce the channel estimation and signal feedback overhead  as in  \cite{havary2008distributed,joint2014cao,Ponukumati2012robust}.}
\begin{align}
R & = {{\mathbb E}_{{\bf{\hat h}}}}\left\{ {\frac{{{T_c} - L}}{{{T_c}}}{{\log }_2}\left( {1 + \frac{{{{\left| {{{{\bf{\hat h}}}^T}{\bf{w}}} \right|}^2}}}{{{\mathbb E}\left\{ {{{\left| {{\bf{h}}_e^T{\bf{w}}} \right|}^2}} \right\} + \sigma _t^2}}} \right)} \right\}\notag\\
& \overset{(a)} {\le} \frac{{{T_c} - L}}{{{T_c}}}{\log _2}\left( {1 + \frac{{{{\bf{w}}^H}{{\bf{R}}_{{\bf{\hat h}}}}{\bf{w}}}}{{{{\bf{w}}^H}{{\bf{R}}_{{{\bf{h}}_e}}}{\bf{w}} + \sigma _t^2}}} \right)\label{comm_rate}\\
& \overset{(b)} {=}\frac{{{T_c} - L}}{{{T_c}}}{\log _2}\left( {\frac{{{{\bf{w}}^H}{{\bf{R}}_{\bf{h}}}{\bf{w}} + \sigma _t^2}}{{{{\bf{w}}^H}{{\bf{R}}_{{{\bf{h}}_e}}}{\bf{w}} + \sigma _t^2}}} \right),
\end{align}
where inequality $(a)$ holds since  $\log \left(  \cdot  \right)$  is a concave function  and equality $(b)$ holds due to identity ${{\bf{R}}_{\bf{h}}} = {{\bf{R}}_{{\bf{\hat h}}}} + {{\bf{R}}_{{{\bf{h}}_e}}}$, in which ${{\bf{R}}_{{{\bf{h}}_e}}}$ is given by 
\begin{align}
\!\!	{{\bf{R}}_{{{\bf{h}}_e}}}{\rm{ = }}{\mathbb E}\left\{ {{\bf{h}}_e^*{\bf{h}}_e^T} \right\}&={{\bf{R}}_{\bf{h}}} - {{\bf{R}}_{\bf{h}}}{\bf{X}}_c^H{\left( {{{\bf{X}}_c}{{\bf{R}}_{\bf{h}}}{\bf{X}}_c^H \!+\! \sigma _p^2{{\bf{I}}_L}} \right)^{ - 1}}{{\bf{X}}_c}{{\bf{R}}_{\bf{h}}}\notag\\
	& = \sigma _p^2{{\bf{R}}_{\bf{h}}}{\left( {{\bf{X}}_c^H{{\bf{X}}_c}{{\bf{R}}_{\bf{h}}} + \sigma _p^2{{\bf{I}}_{{N_t}}}} \right)^{ - 1}}.
\end{align}

\subsection{Problem Formulation}
Our objective  is to  maximize the target detection probability   by jointly optimizing the pilot  matrix, transmit beamformer, and training duration, subject to  the minimum transmission rate required by the user. Accordingly, the problem is formulated as  follows
 \begin{subequations} \label{P}
	\begin{align}
	& \mathop {\max }\limits_{{{\bf{X}}_c},{\bf{w}},L} {{\mathbb E}_{{{\bf{X}}_t},{{\bf{g}}_i}}}\left\{ {{Q_{{N_r}\min \left( {L + 1,{N_t}} \right)}}\left( {\sqrt 2 {{\left\| {\frac{1}{{{\sigma _r}}}{\bf{XG}}} \right\|}_F},\sqrt {2{{\hat \Gamma }_{{\rm{th}}}}} } \right)} \right\}\label{P_obj}\\
	& {\rm s.t.}~ \frac{{{T_c} - L}}{{{T_c}}}{\log _2}\left( {\frac{{{{\bf{w}}^H}{{\bf{R}}_{\bf{h}}}{\bf{w}} + \sigma _t^2}}{{{{\bf{w}}^H}{{\bf{R}}_{{{\bf{h}}_e}}}{\bf{w}} + \sigma _t^2}}} \right) \ge {R_{{\rm{th}}}},\label{P_const1}\\
	& \qquad  \frac{{\left\| {{{\bf{X}}_c}} \right\|_F^2 + \left( {{T_c} - L} \right){{\left\| {\bf{w}} \right\|}^2}}}{{{T_c}}} \le {P_{{\rm{ave}}}},\label{P_const2}\\
			& \qquad 0 \le L \le T_c, L \in {\mathbb Z}^+, \label{P_const3}
	\end{align}
\end{subequations}
where ${R_{{\rm{th}}}}$ in \eqref{P_const1} denotes the minimum transmission rate required by the user and ${P_{\rm ave }}$
in \eqref{P_const2} stands for the average power constraint.\footnote{The peak power constraint for pilot sequence design  is not considered here  since the total energy  is   allocated  for each symbol   either with  a water-filling manner or with an equal allocation manner, which can be clearly seen late  in Section III-A.  This indicates that the peak power will not be significantly large and thus can be ignored here.}
 

Problem \eqref{P} is challenging to solve due to the following reasons: 1) the generalized Marcum  $Q_m(a,b)$ function in \eqref{P_obj} has no explicit expression w.r.t. $a$  and  $b$, which cannot be deterministically  analyzed; 2) the  transmit beamforming vector $\bf w$, the  pilot matrix ${\bf X}_c$, and training duration $L$ are intricately coupled in constraint \eqref{P_const1}; 3) constraint  \eqref{P_const3} involves an integer  variable $L$.  In general,  there are no standard methods for solving such a non-convex optimization problem optimally. Nevertheless, we first unveil the hidden pilot structure by studying the optimal pilot design for both communication and target detection, and then  propose an efficient algorithm to solve problem \eqref{P} in the following section.
\section{Proposed Solution}
\subsection{Information-Theoretic Approach for Pilot Design}
Denote  the MI  between the received signals  ${{\bf{y}}_{r,i}}$'s and the channel reflection coefficients ${\bf g}_i$'s under hypothesis ${\cal H}_1$ by $I\left( {{{\bf{y}}_{r,1}} \ldots ,{{\bf{y}}_{r,{N_r}}};{{\bf{g}}_1}, \ldots ,{{\bf{g}}_{{N_r}}}|{{\cal H}_1}} \right)$, which can be expressed as  
\begin{align}
&I\left( {{{\bf{y}}_{r,1}} \ldots ,{{\bf{y}}_{r,{N_r}}};{{\bf{g}}_1}, \ldots ,{{\bf{g}}_{{N_r}}}|{{\cal H}_1}} \right)=\notag\\
&\qquad\qquad\qquad \qquad \qquad \qquad  {N_r}{\log _2}\left| {{{\bf{I}}_{{T_c}}} + \frac{{\delta _g^2{\bf{X}}{{\bf{X}}^H}}}{{\sigma _r^2}}} \right|.
 \label{mutual_information}
\end{align}
 Since ${\bf X}_t$ is a random variable due to   the random data for transmission,  ${\bf X}$ is  also a random variable. Furthermore, as the
 probability distribution of ${\bf X}$ is difficult to obtain, we are interested in the expected MI, which can be obtained as 
 \begin{align}
 &\hat I\left( {{{\bf{y}}_{r,1}} \ldots ,{{\bf{y}}_{r,{N_r}}};{{\bf{g}}_1}, \ldots ,{{\bf{g}}_{{N_r}}}|{{\cal H}_1}} \right)\notag\\
 &={\mathbb E}\left\{ {I\left( {{{\bf{y}}_{r,1}} \ldots ,{{\bf{y}}_{r,{N_r}}};{{\bf{g}}_1}, \ldots ,{{\bf{g}}_{{N_r}}}|{{\cal H}_1}} \right)} \right\}\notag\\
 &\le {N_r}{\log _2}\left| {{{\bf{I}}_{{T_c}}} + \frac{{\delta _g^2{\mathbb E}\left\{ {{\bf{X}}{{\bf{X}}^H}} \right\}}}{{\sigma _r^2}}} \right|\notag\\
&= {N_r}{\log _2}\left| {{{\bf{I}}_L}{\rm{ + }}\frac{{\delta _g^2}}{{\sigma _r^2}}{{\bf{X}}_c}{\bf{X}}_c^H} \right| + {N_r}{\log _2}\left| {{{\bf{I}}_{{T_c} - L}}{\rm{ + }}\frac{{\delta _g^2}}{{\sigma _r^2}}{{\bf{I}}_{{T_c} - L}}} \right|\notag\\
&={N_r}{\log _2}\left| {{{\bf{I}}_L}{\rm{ + }}\frac{{\delta _g^2}}{{\sigma _r^2}}{{\bf{X}}_c}{\bf{X}}_c^H} \right| + {N_r}\left( {{T_c} - L} \right){\log _2}\left( {1 + \frac{P_t{\delta _g^2}}{{\sigma _r^2}}} \right),
\end{align}
where ${P_t} = {\left\| {\bf{w}} \right\|^2}$.

As such, the corresponding optimization problem can be formulated as 
\begin{subequations} \label{P0}
	\begin{align}
	& \mathop {\max }\limits_{{{\bf{X}}_c},{\bf w},P_t \ge 0,L} {N_r}{\log _2}\left| {{{\bf{I}}_L}+\frac{{\delta _g^2}}{{\sigma _r^2}}{{\bf{X}}_c}{\bf{X}}_c^H} \right| \notag\\
	&\qquad\qquad\qquad\qquad + {N_r}\left( {{T_c} - L} \right){\log _2}\left( {1 + \frac{P_t{\delta _g^2}}{{\sigma _r^2}}} \right)\label{P0_obj}\\
	& {\rm s.t.}~  {\left\| {\bf{w}} \right\|^2} \ge {P_t}, \label{P0const1} \\
	& \qquad \eqref{P_const1}, \eqref{P_const2}, \eqref{P_const3}.
	\end{align}
\end{subequations}
Although the objective function  \eqref{P0_obj}    is much simplified when compared to \eqref{P_obj}, problem \eqref{P0} is still difficult to solve. It should be pointed out that at the optimal solution of  problem \eqref{P0}, the inequality in \eqref{P0const1} must be met with equality.
We note that incorporating ${{\bf{X}}_c^H{{\bf{X}}_c}}$ as one optimization  variable  and then directly optimizing ${{\bf{X}}_c^H{{\bf{X}}_c}}$  fails to work here
 since  ${{\bf{X}}_c^H{{\bf{X}}_c}{{\bf{R}}_{\bf{h}}}}$ is not a Hermitian matrix and  ${\rm{tr}}\left( {{{\bf{R}}_{\bf{h}}}{{\left( {{\bf{X}}_c^H{{\bf{X}}_c}{{\bf{R}}_{\bf{h}}} + \sigma _p^2{{\bf{I}}_L}} \right)}^{ - 1}}} \right)$ in \eqref{P_const1}  is thus not convex w.r.t ${{\bf{X}}_c^H{{\bf{X}}_c}}$ in general. 
In fact,  the key challenge for solving   problem \eqref{P0} lies in optimizing   the pilot matrix ${\bf X}_c$. Motivated by this, we first study the pilot matrix design for   minimizing the channel MSE and maximizing the MI of the target, respectively. Then, a general (and flexible)  pilot matrix  structure for  balancing  the channel estimation and the target detection is proposed.
\subsubsection{Optimal Pilot  Design for Channel Estimation}
In this scenario, we aim to design optimal pilot matrix ${\bf X}_c$ to minimize the MSE of the communication channel $\bf h$.
 The    corresponding optimization problem is formulated as 
 \begin{subequations} \label{OptimalPilotDesign}
	\begin{align}
	&\mathop {\min }\limits_{{{\bf{X}}_c}} \sigma _p^2{\rm{tr}}\left( {{{\bf{R}}_{\bf{h}}}{{\left( {{\bf{X}}_c^H{{\bf{X}}_c}{{\bf{R}}_{\bf{h}}} + \sigma _p^2{{\bf{I}}_L}} \right)}^{ - 1}}} \right) \label{OptimalPilotDesign_obj}\\
	& {\rm s.t.}~{\left\| {{{\bf{X}}_c}} \right\|_F^2} = L.\label{OptimalPilotDesign_const1}
	\end{align}
\end{subequations}
According to \cite[Theorem 1]{Palomar2003joint}, the optimal pilot matrix ${\bf X}_c$ has the form  of 
\begin{align}
{\bf{X}}_c^{H,{\rm{opt}}} = \left\{ {\begin{array}{*{20}{l}}
	{\left[ {\begin{array}{*{20}{c}}
			{{{\bf{U}}_{\bf{h}}}}&{{{\bf{0}}_{{N_t} \times \left( {L - {N_t}} \right)}}}
			\end{array}} \right]{{\bf{\Lambda }}_1},{\kern 1pt} {\kern 1pt} {\kern 1pt} {\kern 1pt} {\kern 1pt} {\rm{if}}{\kern 1pt} {\kern 1pt} {\kern 1pt} {\kern 1pt} {\kern 1pt} L \ge {N_t},}\\
	{{{\bf{U}}_{\bf{h}}}\left( {1:L} \right){{\bf{\Lambda }}_2},{\kern 1pt} {\kern 1pt} {\kern 1pt} {\kern 1pt} {\kern 1pt} {\rm otherwise},}
	\end{array}} \right.
\end{align}
 where  ${{{\bf{U}}_{\bf{h}}}}$ results from the eigendecomposition of  ${{\bf{R}}_{\bf{h}}} = {{\bf{U}}_{\bf{h}}}{{\bf{\Sigma }}_{\bf{h}}}{\bf{U}}_{\bf{h}}^H$,  and ${{\bf{\Lambda }}_1}$ and ${{\bf{\Lambda }}_2}$ represent the diagonal matrix where each diagonal entry is determined as follows. 
 
For the case $L \ge {N_t}$, substituting ${\bf{X}}_c^{H,{\rm{opt}}} = \left[ {\begin{array}{*{20}{c}}
 	{{{\bf{U}}_{\bf{h}}}}&{{{\bf{0}}_{{N_t} \times \left( {L - {N_t}} \right)}}}
 	\end{array}} \right]{{\bf{\Lambda }}_1}$ into 
 \eqref{OptimalPilotDesign_obj}, we have  \eqref{term1} at the top of  next page, where  ${{{\bf{\Sigma }}_{{\bf{h}},i,i}}}$  and ${{\bf{\Lambda }}_{1,i,i}}$  denote the $i$th diagonal entry of ${{{\bf{\Sigma }}_{\bf{h}}}}$ and ${{{\bf{\Lambda }}_1}}$, respectively.
 
 \newcounter{mytempeqncnt0}
 \begin{figure*}
 	\normalsize
 	\setcounter{mytempeqncnt0}{\value{equation}}
 	\begin{align}
 & \sigma _p^2{\rm{tr}}\left( {{{\bf{R}}_{\bf{h}}}{{\left( {{\bf{X}}_c^H{{\bf{X}}_c}{{\bf{R}}_{\bf{h}}} + \sigma _p^2{{\bf{I}}_{N_t}}} \right)}^{ - 1}}} \right)\notag\\
 & = \sigma _p^2{\rm{tr}}\left( {{{\bf{U}}_{\bf{h}}}{{\bf{\Sigma }}_{\bf{h}}}{\bf{U}}_{\bf{h}}^H{{\left( {\left[ {\begin{array}{*{20}{c}}
 					{{{\bf{U}}_{\bf{h}}}}&{{{\bf{0}}_{{N_t} \times \left( {L - {N_t}} \right)}}}
 					\end{array}} \right]{{\bf{\Lambda }}_1}{\bf{\Lambda }}_1^H\left[ {\begin{array}{*{20}{c}}
 					{{\bf{U}}_{\bf{h}}^H}\\
 					{{{\bf{0}}_{\left( {L - {N_t}} \right) \times {N_t}}}}
 					\end{array}} \right]{{\bf{U}}_{\bf{h}}}{{\bf{\Sigma }}_{\bf{h}}}{\bf{U}}_{\bf{h}}^H + \sigma _p^2{{\bf{I}}_{N_t}}} \right)}^{ - 1}}} \right)\notag\\
 & = \sigma _p^2{\rm{tr}}\left( {{{\bf{\Sigma }}_{\bf{h}}}{{\left( {\left[ {\begin{array}{*{20}{c}}
 					{{{\bf{I}}_{{N_t}}}}&{{{\bf{0}}_{{N_t} \times \left( {L - {N_t}} \right)}}}
 					\end{array}} \right]{{\bf{\Lambda }}_1}{\bf{\Lambda }}_1^H\left[ {\begin{array}{*{20}{c}}
 					{{{\bf{I}}_{{N_t}}}}\\
 					{{{\bf{0}}_{\left( {L - {N_t}} \right) \times {N_t}}}}
 					\end{array}} \right]{{\bf{\Sigma }}_{\bf{h}}} + \sigma _p^2{{\bf{I}}_{N_t}}} \right)}^{ - 1}}} \right)		\notag\\
 & = \sum\limits_{i = 1}^{{N_t}} {\frac{{{{\bf{\Sigma }}_{{\bf{h}},i,i}}\sigma _p^2}}{{{\bf{\Lambda }}_{1,i,i}^2{{\bf{\Sigma }}_{{\bf{h}},i,i}} + \sigma _p^2}}}. \label{term1}
 	\end{align}
 	\hrulefill 
 \end{figure*}

In addition, plugging ${\bf{X}}_c^{H{\rm{,opt}}} = \left[ {{{\bf{U}}_{\bf{h}}}{\kern 1pt} {\kern 1pt} {\kern 1pt} {\kern 1pt} {{\bf{0}}_{{N_t} \times \left( {L - {N_t}} \right)}} } \right]{{\bf{\Lambda }}_1}$ into  \eqref{OptimalPilotDesign_const1} yields 
\begin{align}
\sum\limits_{i = 1}^{{N_t}} {{\bf{\Lambda }}_{1,i,i}^2 = L}. 
\end{align} 
 As a result, problem \eqref{OptimalPilotDesign} is simplified as 
  \begin{subequations} \label{OptimalPilotDesign_new}
 	\begin{align}
 	&\mathop {\min }\limits_{{\bf{\Lambda }}_{1,i,i}^{}} \sum\limits_{i = 1}^{{N_t}} {\frac{{{{\bf{\Sigma }}_{{\bf{h}},i,i}}\sigma _p^2}}{{{\bf{\Lambda }}_{1,i,i}^2{{\bf{\Sigma }}_{{\bf{h}},i,i}} + \sigma _p^2}}}  \label{OptimalPilotDesign_new_obj}\\
 	& {\rm s.t.}~\sum\limits_{i = 1}^{{N_t}} {{\bf{\Lambda }}_{1,i,i}^2 \le  L}.\label{OptimalPilotDesign_new_const1}
 	\end{align}
 \end{subequations}
Note that at the optimal solution, constraint \eqref{OptimalPilotDesign_new_const1}    must be met with equality.
We can readily verify that problem \eqref{OptimalPilotDesign_new} is a convex optimization problem, for which we can apply the  Lagrange duality to obtain a semi-closed form  optimal solution given by 
\begin{align}
{\bf{\Lambda }}_{1,i,i}^{{\rm{opt}}} = \left\{ \begin{array}{l}
\sqrt {\mu_1  - \frac{{\sigma _p^2}}{{{{\bf{\Sigma }}_{{\bf{h}},i,i}}}}} ,{\kern 1pt} {\kern 1pt} {\kern 1pt} {\kern 1pt} {\kern 1pt} {\kern 1pt} {\kern 1pt} {\kern 1pt} {\kern 1pt} {\kern 1pt} {\rm{if}}{\kern 1pt} \mu_1  > \frac{{\sigma _p^2}}{{{{\bf{\Sigma }}_{{\bf{h}},i,i}}}}{\kern 1pt} {\kern 1pt} {\kern 1pt} {\rm{and}}{\kern 1pt} {\kern 1pt} {\kern 1pt} {\kern 1pt} {{\bf{\Sigma }}_{{\bf{h}},i,i}} \ne 0,\\
0,{\kern 1pt} {\kern 1pt} {\kern 1pt} {\kern 1pt} {\kern 1pt} {\kern 1pt} {\kern 1pt} {\rm{otherwise}},
\end{array} \right. 
\end{align}
where $\mu_1 $ is  determined by satisfying the power constraint, i.e., $\sum\limits_{i = 1}^{{N_t}} {{{\left( {{\bf{\Lambda }}_{1,i,i}^{{\rm{opt}}}} \right)}^2} = L} $.

For the  case $L < {N_t}$, substituting ${\bf{X}}_c^{H{\rm{,opt}}} = {{\bf{U}}_{\bf{h}}}\left( {1:L} \right){{\bf{\Lambda }}_2}$
into \eqref{OptimalPilotDesign_obj} yields \eqref{chanel_estimiation_pilot} at the top of this page, where ${{\bf{\Lambda }}_{2,i,i}}$  denotes the $i$th diagonal entry of  ${{{\bf{\Lambda }}_2}}$.

 \newcounter{mytempeqncnt1}
\begin{figure*}
	\normalsize
	\setcounter{mytempeqncnt1}{\value{equation}}
	\begin{align}
	& \sigma _p^2{\rm{tr}}\left( {{{\bf{R}}_{\bf{h}}}{{\left( {{\bf{X}}_c^H{{\bf{X}}_c}{{\bf{R}}_{\bf{h}}} + \sigma _p^2{{\bf{I}}_{N_t}}} \right)}^{ - 1}}} \right)=
\sigma _p^2{\rm{tr}}\left( {{{\bf{U}}_{\bf{h}}}{{\bf{\Sigma }}_{\bf{h}}}{\bf{U}}_{\bf{h}}^H{{\left( {{{\bf{U}}_{\bf{h}}}\left( {1:L} \right){{\bf{\Lambda }}_2}{\bf{\Lambda }}_2^H{\bf{U}}_{\bf{h}}^H\left( {1:L} \right){{\bf{U}}_{\bf{h}}}{{\bf{\Sigma }}_{\bf{h}}}{\bf{U}}_{\bf{h}}^H + \sigma _p^2{{\bf{I}}_{N_t}}} \right)}^{ - 1}}} \right)\notag\\
&   = \sigma _p^2{\rm{tr}}\left( {{{\bf{\Sigma }}_{\bf{h}}}{{\left( {\left[ {\begin{array}{*{20}{c}}
					{{{\bf{I}}_L}}\\
					{{{\bf{0}}_{\left( {{N_t} - L} \right) \times L}}}
					\end{array}} \right]{{\bf{\Lambda }}_2}{\bf{\Lambda }}_2^H\left[ {\begin{array}{*{20}{c}}
					{{{\bf{I}}_L}}&{{{\bf{0}}_{L \times \left( {{N_t} - L} \right)}}}
					\end{array}} \right]{{\bf{\Sigma }}_{\bf{h}}} + \sigma _p^2{{\bf{I}}_{N_t}}} \right)}^{ - 1}}} \right)	\notag\\
& = \sum\limits_{i = 1}^L {\frac{{{{\bf{\Sigma }}_{{\bf{h}},i,i}}\sigma _p^2}}{{{\bf{\Lambda }}_{2,i,i}^2{{\bf{\Sigma }}_{{\bf{h}},i,i}} + \sigma _p^2}}} {\rm{ + }}\sum\limits_{i = L+1}^{{N_t}} {{{\bf{\Sigma }}_{{\bf{h}},i,i}}} . \label{chanel_estimiation_pilot}
	\end{align}
	\hrulefill 
\end{figure*}

Therefore,  problem \eqref{OptimalPilotDesign} is simplified as 
  \begin{subequations} \label{OptimalPilotDesign_new_1}
	\begin{align}
		&\mathop {\min }\limits_{{\bf{\Lambda }}_{2,i,i}^{}} \sum\limits_{i = 1}^L {\frac{{{{\bf{\Sigma }}_{{\bf{h}},i,i}}\sigma _p^2}}{{{\bf{\Lambda }}_{2,i,i}^2{{\bf{\Sigma }}_{{\bf{h}},i,i}} + \sigma _p^2}}{\rm{ + }}\sum\limits_{i =L+ 1}^{{N_t}} {{{\bf{\Sigma }}_{{\bf{h}},i,i}}} }     \label{OptimalPilotDesign_new_1_obj}\\
		& {\rm s.t.}~\sum\limits_{i = 1}^{{L}} {{\bf{\Lambda }}_{2,i,i}^2 \le  L}.\label{OptimalPilotDesign_new_1_const1}
	\end{align}
\end{subequations}
Similar to the case  $L\ge N_t$, the optimal semi-closed form solution for the case $L<N_t$  can be obtained by using the Lagrange duality, which is given by   
\begin{align}
	{\bf{\Lambda }}_{2,i,i}^{{\rm{opt}}} = \left\{ \begin{array}{l}
		\sqrt {\mu_2  - \frac{{\sigma _p^2}}{{{{\bf{\Sigma }}_{{\bf{h}},i,i}}}}} ,{\kern 1pt} {\kern 1pt} {\kern 1pt} {\kern 1pt} {\kern 1pt} {\kern 1pt} {\kern 1pt} {\kern 1pt} {\kern 1pt} {\kern 1pt} {\rm{if}}{\kern 1pt} \mu_2  > \frac{{\sigma _p^2}}{{{{\bf{\Sigma }}_{{\bf{h}},i,i}}}}{\kern 1pt} {\kern 1pt} {\kern 1pt} {\rm{and}}{\kern 1pt} {\kern 1pt} {\kern 1pt} {\kern 1pt} {{\bf{\Sigma }}_{{\bf{h}},i,i}} \ne 0, \\
		0,{\kern 1pt} {\kern 1pt} {\kern 1pt} {\kern 1pt} {\kern 1pt} {\kern 1pt} {\kern 1pt} {\rm{otherwise}},
	\end{array} \right. \label{optimal_solution_Lba}
\end{align}
where $\mu_2 $ is  determined by satisfying the power constraint, i.e., $\sum\limits_{i = 1}^{{L}} {{{\left( {{\bf{\Lambda }}_{2,i,i}^{{\rm{opt}}}} \right)}^2} = L}$.  

Therefore, the optimal covariance matrix ${\bf{X}}_c^{{\rm{opt}}}{\bf{X}}_c^{H,{\rm{opt}}}$ satisfies  
\begin{align}
&{\bf{X}}_c^{{\rm{opt}}}{\bf{X}}_c^{H,{\rm{opt}}} = \notag\\
&\left\{ \begin{array}{l}
{\bf{\Lambda }}_1^{H,{\rm{opt}}}\left[ {\begin{array}{*{20}{c}}
	{{{\bf{I}}_{{N_t}}}}&{{{\bf{0}}_{{N_t} \times \left( {L - {N_t}} \right)}}}\\
	{{{\bf{0}}_{\left( {L - {N_t}} \right) \times {N_t}}}}&{{{\bf{0}}_{L - {N_t}}}}
	\end{array}} \right]{\bf{\Lambda }}_1^{{\rm{opt}}},{\kern 1pt} {\kern 1pt} {\kern 1pt} {\kern 1pt} {\kern 1pt} {\rm{if}}{\kern 1pt} {\kern 1pt} {\kern 1pt} {\kern 1pt} {\kern 1pt} L \ge {N_t},\\
{\bf{\Lambda }}_2^{H,{\rm{opt}}}{\bf{\Lambda }}_2^{{\rm{opt}}},{\kern 1pt} {\kern 1pt} {\kern 1pt} {\kern 1pt} {\kern 1pt} {\rm otherwise}.
\end{array} \right. \label{optimal_commun_pilot}
\end{align}
Note that if the communication channel $\bf h$ is an uncorrelated Rayleigh fading channel, its covariance matrix is reduced to ${{\bf{R}}_{\bf{h}}} = \delta _h^2{{\bf{I}}_{{N_t}}}$, where $\delta _h^2$ represents the channel power gain. We thus have ${\bf{\Lambda }}_{1,1,1}^{{\rm{opt}}} =  \ldots= {\bf{\Lambda }}_{1,{N_t},{N_t}}^{{\rm{opt}}} = \sqrt {L/{N_t}} $ and ${\bf{\Lambda }}_{2,1,1}^{{\rm{opt}}} =  \ldots={\bf{\Lambda }}_{2,L,L}^{{\rm{opt}}} = 1$.
 As a result,  \eqref{optimal_commun_pilot} is reduced to 
\begin{align}
&{\bf{X}}_c^{{\rm{opt}}}{\bf{X}}_c^{H,{\rm{opt}}} = \notag\\
&\left\{ \begin{array}{l}
\frac{L}{{{N_t}}}\left[ {\begin{array}{*{20}{c}}
	{{{\bf{I}}_{{N_t}}}}&{{{\bf{0}}_{{N_t} \times \left( {L - {N_t}} \right)}}}\\
	{{{\bf{0}}_{\left( {L - {N_t}} \right) \times {N_t}}}}&{{{\bf{0}}_{L - {N_t}}}}
	\end{array}} \right],{\kern 1pt} {\kern 1pt} {\kern 1pt} {\kern 1pt} {\kern 1pt} {\rm{if}}{\kern 1pt} {\kern 1pt} {\kern 1pt} {\kern 1pt} {\kern 1pt} L \ge {N_t},\\
{{\bf{I}}_L},{\kern 1pt} {\kern 1pt} {\kern 1pt} {\kern 1pt} {\kern 1pt} {\rm otherwise}.
\end{array} \right.\label{optimal_commun_pilot_1}
\end{align}

\textbf{\emph{Remark 1:}} 
Based on  \eqref{OptimalPilotDesign_new} and \eqref{OptimalPilotDesign_new_1}, we observe  that  a larger $L$ leads to a smaller objective value, which indicates that a larger length of pilot sequences is beneficial for reducing the channel estimation error. The reason is that the available power in constraints \eqref{OptimalPilotDesign_new_const1} and \eqref{OptimalPilotDesign_new_1_const1}  increases as $L$ increases. While the available power is fixed, namely $\sum\limits_{i = 1}^{L{\kern 1pt} \left( {{N_t}} \right)} {{\bf{\Lambda }}_{2,i,i}^2 \le \hat L} $ with fixed ${\hat L}$, we can readily verify that increasing $L$ will not  improve the  channel estimation accuracy for the case of  $L > {\rm{rank}}\left( {{{\bf{R}}_{\bf{h}}}} \right)$.

\subsubsection{Pilot  Design for Target Detection}
 The    corresponding optimization problem is formulated as 
\begin{subequations} \label{PilotDesign_Estimation}
	\begin{align}
	&\!\!\mathop {\max }\limits_{{{\bf{X}}_c}} {N_r}{\log _2}\left| {{{\bf{I}}_L}{\rm{ + }}\frac{{\delta _g^2}}{{\sigma _r^2}}{{\bf{X}}_c}{\bf{X}}_c^H} \right| + {N_r}\left( {{T_c} - L} \right){\log _2}\left( {1 + \frac{P_t{\delta _g^2}}{{\sigma _r^2}}} \right)\label{P_case1}\\
	& {\rm s.t.}~{\left\| {{{\bf{X}}_c}} \right\|_F^2} = L.
	\end{align}
\end{subequations}
Since ${\bf{X}}_c^H{{\bf{X}}_c}$ is a positive semi-definite matrix, its eigen-decomposition can be expressed as ${\bf{X}}_c^H{{\bf{X}}_c} = {{\bf{U}}_c}{{\bf{\Sigma }}_c}{\bf{U}}_c^H$. Substituting it into ${\log _2}\left| {{{\bf{I}}_L}{\rm{ + }}\frac{{\delta _g^2}}{{\sigma _r^2}}{{\bf{X}}_c}{\bf{X}}_c^H} \right|$ yields 
\begin{align}
{\log _2}\left| {{{\bf{I}}_L}{\rm{ + }}\frac{{\delta _g^2}}{{\sigma _r^2}}{{\bf{X}}_c}{\bf{X}}_c^H} \right|& = {\log _2}\left| {{{\bf{I}}_{{N_t}}}{\rm{ + }}\frac{{\delta _g^2}}{{\sigma _r^2}}{\bf{X}}_c^H{{\bf{X}}_c}} \right| \notag\\
&= \sum\limits_{i = 1}^{\min \left( {L,{N_t}} \right)} {{{\log }_2}\left( {1{\rm{ + }}\frac{{\delta _g^2}}{{\sigma _r^2}}{{\bf{\Sigma }}_{c,i,i}}} \right)} , \label{mu_information}
\end{align}
where ${{{\bf{\Sigma }}_{c,i,i}}}$ accounts for the $i$th diagonal entry of ${{{\bf{\Sigma }}_c}}$. Since ${\log _2}\left( {1{\rm{ + }}\frac{{\delta _g^2}}{{\sigma _r^2}}{{\bf{\Sigma }}_{c,i,i}}} \right)$  is a concave function of ${{{\bf{\Sigma }}_{c,i,i}}}$,   we can  
apply the Jensen's inequality to obtain the upper bound of  \eqref{mu_information}  as 
\begin{align}
&\sum\limits_{i = 1}^{\min \left( {L,{N_t}} \right)} {{{\log }_2}\left( {1{\rm{ + }}\frac{{\delta _g^2}}{{\sigma _r^2}}{{\bf{\Sigma }}_{c,i,i}}} \right)}  \le \notag\\
&\qquad\qquad\min \left( {L,{N_t}} \right){\log _2}\left( {1{\rm{ + }}\frac{{\delta _g^2\sum\limits_{i = 1}^{\min \left( {L,{N_t}} \right)} {{{\bf{\Sigma }}_{c,i,i}}} }}{{\sigma _r^2\min \left( {L,{N_t}} \right)}}} \right), \label{mu_information_1}
\end{align}
where equality holds if and only if ${{\bf{\Sigma }}_{c,1,1}} =  \ldots  = {{\bf{\Sigma }}_{c,\min \left( {L,{N_t}} \right),\min \left( {L,{N_t}} \right)}}$. Recall that ${{\bf{X}}_c}{\bf{X}}_c^H = {{\bf{U}}_c}{{\bf{\Sigma }}_c}{\bf{U}}_c^H$. Then, we can  rewrite $\left\| {{{\bf{X}}_c}} \right\|_F^2$ as 
\begin{align}
\left\| {{{\bf{X}}_c}} \right\|_F^2 = L \Rightarrow {\rm{tr}}\left( {{{\bf{U}}_c}{{\bf{\Sigma }}_c}{\bf{U}}_c^H} \right) = L \Rightarrow \sum\limits_{i = 1}^{\min \left( {L,{N_t}} \right)} {{{\bf{\Sigma }}_{c,i,i}}}  = L. \label{mu_information_2}
\end{align}
Therefore, the optimal  pilot matrix  is given by  
\begin{align}
{\bf{X}}_c^{H,{\rm{opt}}} = \left\{ {\begin{array}{*{20}{l}}
	{\sqrt {\frac{L}{{{N_t}}}} \left[ {{{\bf{U}}_c}{\kern 1pt} {\kern 1pt} {\kern 1pt} {\kern 1pt}{\kern 1pt} {\kern 1pt}{{\bf{0}}_{{N_t} \times \left( {L - {N_t}} \right)}}{\kern 1pt} } \right],{\kern 1pt} {\kern 1pt} {\kern 1pt} {\kern 1pt} {\kern 1pt} {\kern 1pt} {\rm{if}}{\kern 1pt} {\kern 1pt} {\kern 1pt} {\kern 1pt} L \ge {N_t},}\\
	{{{\bf{U}}_c}\left( {1:L} \right),{\kern 1pt} {\kern 1pt} {\kern 1pt} {\kern 1pt} {\kern 1pt} {\kern 1pt} {\rm otherwise},}
	\end{array}} \right.
\end{align}
which indicates that  
\begin{align}
&{\bf{X}}_c^{{\rm{opt}}}{\bf{X}}_c^{H,{\rm{opt}}} =\notag\\
& \left\{ \begin{array}{l}
\frac{L}{{{N_t}}}\left[ {\begin{array}{*{20}{c}}
	{{{\bf{I}}_{{N_t}}}}&{{{\bf{0}}_{{N_t} \times \left( {L - {N_t}} \right)}}}\\
	{{{\bf{0}}_{\left( {L - {N_t}} \right) \times {N_t}}}}&{{{\bf{0}}_{  L-{N_t}}}}
	\end{array}} \right],{\kern 1pt} {\kern 1pt} {\kern 1pt} {\kern 1pt} {\kern 1pt} {\rm{if}}{\kern 1pt} {\kern 1pt} {\kern 1pt} L \ge {N_t},\\
{{\bf{I}}_L},{\kern 1pt} {\kern 1pt} {\kern 1pt} {\kern 1pt} {\kern 1pt} {\kern 1pt}  {\rm otherwise}.
\end{array} \right.\label{optimal_detection_pilot}
\end{align}
It is observed  that each pilot sequence should be orthogonal and the optimal  pilot matrix  to problem \eqref{PilotDesign_Estimation} is a unitary-type matrix. 

\textbf{\emph{Remark 2:}} 
Based on \eqref{mu_information_1} and \eqref{mu_information_2}, we can obtain the maximum value of ${\log _2}\left| {{{\bf{I}}_L}{\rm{ + }}\frac{{\delta _g^2}}{{\sigma _r^2}}{{\bf{X}}_c}{\bf{X}}_c^H} \right|$ given by  
$\min \left( {L,{N_t}} \right){\log _2}\left( {1{\rm{ + }}\frac{{\delta _g^2L}}{{\sigma _r^2\min \left( {L,{N_t}} \right)}}} \right)$. It can be readily verified that as $L$ increases, the value of ${\log _2}\left| {{{\bf{I}}_L}{\rm{ + }}\frac{{\delta _g^2}}{{\sigma _r^2}}{{\bf{X}}_c}{\bf{X}}_c^H} \right|$ increases even when $L\ge N_t$. This is because   the available power is proportional to $L$, i.e., ${\left\| {{{\bf{X}}_c}} \right\|_F^2} = L$. It is worth pointing out that in the case with fixed total power, say $\hat L$,  the conclusion will be different. Specifically, the maximum value of ${\log _2}\left| {{{\bf{I}}_L}{\rm{ + }}\frac{{\delta _g^2}}{{\sigma _r^2}}{{\bf{X}}_c}{\bf{X}}_c^H} \right|$ is given by $\min \left( {L,{N_t}} \right){\log _2}\left( {1{\rm{ + }}\frac{{\delta _g^2\hat L}}{{\sigma _r^2\min \left( {L,{N_t}} \right)}}} \right)$,  and   a larger $L$ will lead to a larger value of ${\log _2}\left| {{{\bf{I}}_L}{\rm{ + }}\frac{{\delta _g^2}}{{\sigma _r^2}}{{\bf{X}}_c}{\bf{X}}_c^H} \right|$ as $L< N_t$, while a larger $L$ beyond the value of $N_t$ does not lead to a larger ${\log _2}\left| {{{\bf{I}}_L}{\rm{ + }}\frac{{\delta _g^2}}{{\sigma _r^2}}{{\bf{X}}_c}{\bf{X}}_c^H} \right|$. 

\textbf{\emph{Remark 3:}} Based on  \eqref{optimal_commun_pilot_1}  and  \eqref{optimal_detection_pilot}, we can find   that the nonunitary pilot training  with unequal power allocation is optimal for  channel estimation to adapt to the communication channel property, while the unitary pilot training  with equal  power allocation  is optimal for target detection. Interestingly,
if the communication channel is a  Rayleigh fading channel, the unitary pilot training   with equal power is optimal for both channel estimation and target detection. 
\subsection{Nonunitary Pilot Matrix-based Algorithm Design}
Since the unitary pilot training  with equal  power can be treated as a special case of the nonunitary pilot training  by setting ${{\bf{\Lambda }}_1}$ and  ${{\bf{\Lambda }}_2}$  as identity matrices, we can adopt the  nonunitary pilot matrix structure as the desired pilot matrix and optimize the power allocation in each  training sequence to maximize the system  utility.
 In the sequel, we next only consider the case of $L<N_t$.  The case of $L\ge N_t$  can be studied similarly, which is thus omitted  for brevity. 

To be specific, for the nonunitary pilot matrix, we set ${\bf{X}}_c^H = {{\bf{U}}_{\bf{h}}}\left( {1:L} \right){\bf{\Lambda }}$, where ${\bf{\Lambda }} = {\rm{diag}}\left( {{{\bf{\Lambda }}_{1,1}}, \ldots ,{{\bf{\Lambda }}_{L,L}}} \right)$ with ${{\bf{\Lambda }}_{i,i}}\ge 0$ needs to be optimized. 
Then, the power of the communication channel estimation error, i.e., ${\sigma _p^2{{\bf{w}}^H}{{\bf{R}}_{\bf{h}}}{{\left( {{\bf{X}}_c^H{{\bf{X}}_c}{{\bf{R}}_{\bf{h}}} + \sigma _p^2{{\bf{I}}_{{N_t}}}} \right)}^{ - 1}}{\bf{w}}}$,
 can be transformed  as
\begin{align}
&\sigma _p^2{{\bf{w}}^H}{{\bf{R}}_{\bf{h}}}{\left( {{\bf{X}}_c^H{{\bf{X}}_c}{{\bf{R}}_{\bf{h}}} + \sigma _p^2{{\bf{I}}_{{N_t}}}} \right)^{ - 1}}{\bf{w}}\overset{\triangle}{=}{\rm{MSE}}\left( {{{\bf{\Lambda }}_{i,i}},{\bf{w}}} \right)\notag\\
& = \sigma _p^2{{\bf{w}}^H}{{\bf{U}}_{\bf{h}}}{{\bf{\Sigma }}_{\bf{h}}}{\bf{U}}_{\bf{h}}^H\left( {{{\bf{U}}_{\bf{h}}}\left( {1:L} \right){\bf{\Lambda }}{{\bf{\Lambda }}^H}{\bf{U}}_{\bf{h}}^H} \right.\left( {1:L} \right){{\bf{U}}_{\bf{h}}}{{\bf{\Sigma }}_{\bf{h}}}\notag\\
&\quad\times{\left. {{\bf{U}}_{\bf{h}}^H + \sigma _p^2{{\bf{I}}_{{N_t}}}} \right)^{ - 1}}{\bf{w}}\notag\\
&=\sigma _p^2{{\bf{w}}^H}{{\bf{U}}_{\bf{h}}}{\bf{\Sigma }}_{\bf{h}}^{\frac{1}{2}}{\left( {{\bf{\Sigma }}_{\bf{h}}^{\frac{1}{2}}{{\bf{R}}_{\bf{\Lambda }}}{\bf{\Sigma }}_{\bf{h}}^{\frac{1}{2}} + \sigma _p^2{{\bf{I}}_{{N_t}}}} \right)^{ - 1}}{\bf{\Sigma }}_{\bf{h}}^{\frac{1}{2}}{\bf{U}}_{\bf{h}}^H{\bf{w}}, \label{nonunitary_1}	
\end{align}
where ${{\bf{R}}_{\bf{\Lambda }}} = \left[ {\begin{array}{*{20}{c}}
	{{\bf{\Lambda }}{{\bf{\Lambda }}^H}}&{{{\bf{0}}_{L \times \left( {{N_t} - L} \right)}}}\\
	{{{\bf{0}}_{\left( {{N_t} - L} \right) \times L}}}&{{{\bf{0}}_{{N_t} - L}}}
	\end{array}} \right]$.

In addition, similar to  \eqref{mu_information},   ${\log _2}\left| {{{\bf{I}}_L}{\rm{ + }}\frac{{\delta _g^2}}{{\sigma _r^2}}{{\bf{X}}_c}{\bf{X}}_c^H} \right|$ can be simplified as 
\begin{align}
{\log _2}\left| {{{\bf{I}}_L}{\rm{ + }}\frac{{\delta _g^2}}{{\sigma _r^2}}{{\bf{X}}_c}{\bf{X}}_c^H} \right| = \sum\limits_{i = 1}^L {{{\log }_2}\left( {1 + \frac{{\delta _g^2{\bf{\Lambda }}_{i,i}^2}}{{\sigma _r^2}}} \right)}.\label{nonunitary_2}	
\end{align}
As a result, based on \eqref{nonunitary_1}  and \eqref{nonunitary_2}, problem \eqref{P0}  is reduced to 
\begin{subequations} \label{P0_nonunitary}
	\begin{align}
	& \mathop {\max }\limits_{{{\bf{\Lambda }}_{i,i}} \ge 0, {\bf w}, P_t \ge 0, L} {N_r}\sum\limits_{i = 1}^L {{{\log }_2}\left( {1 + \frac{{\delta _g^2{\bf{\Lambda }}_{i,i}^2}}{{\sigma _r^2}}} \right)}   \notag\\
	&\qquad \qquad  \qquad  +{N_r}\left( {{T_c} - L} \right){\log _2}\left( {1 + \frac{P_t{\delta _g^2}}{{\sigma _r^2}}} \right)\label{P0_nonunitary_obj}\\
	& {\rm s.t.}~\frac{{{T_c} - L}}{{{T_c}}}{\log _2}\left( {\frac{{{{\bf{w}}^H}{{\bf{R}}_{\bf{h}}}{\bf{w}} + \sigma _t^2}}{{{\rm{MSE}}\left( {{{\bf{\Lambda }}_{i,i}},{\bf{w}}} \right) + \sigma _t^2}}} \right) \ge {R_{{\rm{th}}}}, \label{P0_nonunitary_const1}\\
	& \qquad \sum\limits_{i = 1}^L {{\bf{\Lambda }}_{i,i}^2}  + \left( {{T_c} - L} \right){\left\| {\bf{w}} \right\|^2} \le T_c{P_{\rm  ave }},\label{P0_nonunitary_const2}\\
	& \qquad \eqref{P_const3},\eqref{P0const1}.
	\end{align}
\end{subequations}
We note that the traditional method that relaxes the integer variable $L$ into a continuous variable is not feasible here. Considering  the fact
that $L$ is not large in general due to the limited channel coherence time,   a brutal-force  search is employed to pick up a best solution  from $T_c$ choices with affordable computational
complexity. To be specific, with the fixed $L$, we divide all the optimization variables into two
blocks, namely 1) transmit beamforming vector ${\bf{w}}$ and 2) pilot matrix ${\bf{\Lambda }}$, and then optimize each block in an iterative way, until convergence is achieved.
\subsubsection{Transmit Beamforming Optimization} For a given  pilot matrix  ${{{\bf{\Lambda }}}}$, the subproblem regarding transmit beamforming vector $\bf w$ is given by
\begin{subequations} \label{subproblem1}
	\begin{align}
	& \mathop {\max }\limits_{{\bf{w}},{P_t} \ge 0} {N_r}\sum\limits_{i = 1}^L {{{\log }_2}\left( {1 + \frac{{\delta _g^2{\bf{\Lambda }}_{i,i}^2}}{{\sigma _r^2}}} \right)} \notag\\
	& \qquad \quad + {N_r}\left( {{T_c} - L} \right){\log _2}\left( {1 + \frac{{{P_t}\delta _g^2}}{{\sigma _r^2}}} \right)\label{subproblem1_obj}\\
	& {\rm s.t.}~\eqref{P0const1},\eqref{P0_nonunitary_const1}, \eqref{P0_nonunitary_const2}.
	\end{align}
\end{subequations}
It can be readily verified that the objective function  and constraint \eqref{P0_nonunitary_const2} are convex, while constraints \eqref{P0const1} and  \eqref{P0_nonunitary_const1} are non-convex. However, the left-hand side of \eqref{P0const1} is a convex quadratic function of $\bf w$.  Recall that any convex function is
globally lower-bounded by its first-order Taylor expansion at
any feasible point.  As a result, the successive convex approximation (SCA) technique is
applied. Specifically, for any local point ${\bf w}^r$ at the $r$th iteration, we have
\begin{align}
{\left\| {\bf{w}} \right\|^2} \ge -{\left\| {{{\bf{w}}^r}} \right\|^2} + 2{\mathop{\rm Re}\nolimits} \left\{ {{{\bf{w}}^{r,H}}{\bf{w}}} \right\}\overset{\triangle}{=} {g^{{\rm{lb}}}}\left( {\bf{w}} \right),
\end{align}
where the equality holds at the point ${{\bf{w}}^r}$. Then,  \eqref{P0const1} can be approximated by 
\begin{align}
{g^{{\rm{lb}}}}\left( {\bf{w}} \right) \ge {P_t}, \label{Pt_low}
\end{align}
which is convex since ${g^{{\rm{lb}}}}\left( {\bf{w}} \right)$ is linear  w.r.t. $\bf w$.

To handle the non-convex constraint \eqref{P0_nonunitary_const1},   we first  rewrite it  as 
\begin{align}
{{\bf{w}}^H}{{\bf{R}}_{{\bf{ h}}}}{\bf{w}} &\ge {2^{\frac{{{R_{{\rm{th}}}}{T_c}}}{{{T_c} - L}}}}\left( {\sigma _p^2{{\bf{w}}^H}{{\bf{U}}_{\bf{h}}}{\bf{\Sigma }}_{\bf{h}}^{\frac{1}{2}}{{\left( {{\bf{\Sigma }}_{\bf{h}}^{\frac{1}{2}}{{\bf{R}}_{\bf{\Lambda }}}{\bf{\Sigma }}_{\bf{h}}^{\frac{1}{2}} + \sigma _p^2{{\bf{I}}_{{N_t}}}} \right)}^{ - 1}}} \right.\notag\\
& \times{{\bf{\Sigma }}_{\bf{h}}^{\frac{1}{2}}{\bf{U}}_{\bf{h}}^H{\bf{w}} + \sigma _t^2} \bigg) - \sigma _t^2.
 \label{P0_nonunitary_new_const1_v1}
\end{align}
Since ${{\bf{R}}_{\bf{h}}}$ is positive semi-definite,   ${{\bf{w}}^H}{{\bf{R}}_{\bf{h}}}{\bf{w}}$ is a convex quadratic function of ${\bf{w}}$. Similar to the way for handling constraint \eqref{P0const1},  the SCA  technique is also
applied. 
 Specifically, for any  point ${\bf w}^r$ at the $r$th iteration, we have
\begin{align}
{{\bf{w}}^H}{{\bf{R}}_{\bf{h}}}{\bf{w}} &\ge  - {{\bf{w}}^{r,H}}{{\bf{R}}_{\bf{h}}}{{\bf{w}}^r} + 2{\mathop{\rm Re}\nolimits} \left\{ {{{\bf{w}}^{r,H}}{{\bf{R}}_{\bf{h}}}{\bf{w}}} \right\}\notag\\
&\overset{\triangle}{=} {f^{{\rm{lb}}}}\left( {\bf{w}} \right).
\end{align}
Then, \eqref{P0_nonunitary_new_const1_v1} can be approximated by 
\begin{align}
	{f^{{\rm{lb}}}}\left( {\bf{w}} \right) &\ge  {2^{\frac{{{R_{{\rm{th}}}}{T_c}}}{{{T_c} - L}}}}\left( {\sigma _p^2{{\bf{w}}^H}{{\bf{U}}_{\bf{h}}}{\bf{\Sigma }}_{\bf{h}}^{\frac{1}{2}}{{\left( {{\bf{\Sigma }}_{\bf{h}}^{\frac{1}{2}}{{\bf{R}}_{\bf{\Lambda }}}{\bf{\Sigma }}_{\bf{h}}^{\frac{1}{2}} + \sigma _p^2{{\bf{I}}_{{N_t}}}} \right)}^{ - 1}}} \right.\notag\\
	& \times{{\bf{\Sigma }}_{\bf{h}}^{\frac{1}{2}}{\bf{U}}_{\bf{h}}^H{\bf{w}} + \sigma _t^2} \bigg) - \sigma _t^2, \label{subproblem1_const2}
\end{align}
which is convex since ${f^{{\rm{lb}}}}\left( {\bf{w}} \right)$ is linear  w.r.t. $\bf w$.


Based on  \eqref{Pt_low} and \eqref{subproblem1_const2}, problem \eqref{subproblem1} is transformed into the following problem
\begin{subequations} \label{subproblem1_new}
	\begin{align}
	& \mathop {\max }\limits_{{\bf{w}},{P_t} \ge 0} {N_r}\sum\limits_{i = 1}^L {{{\log }_2}\left( {1 + \frac{{\delta _g^2{\bf{\Lambda }}_{i,i}^2}}{{\sigma _r^2}}} \right)}  \notag\\
	&\qquad \quad + {N_r}\left( {{T_c} - L} \right){\log _2}\left( {1 + \frac{{{P_t}\delta _g^2}}{{\sigma _r^2}}} \right)\\
	& {\rm s.t.}~ \eqref{P0_nonunitary_const2}, \eqref{Pt_low}, \eqref{subproblem1_const2}.
	\end{align}
\end{subequations}
 It can be readily verified that problem \eqref{subproblem1_new} is convex, which can be efficiently solved by convex optimization solvers.

\subsubsection{Pilot Matrix Optimization} Define ${{{\bf{\bar \Lambda }}}_{i,i}} = {\bf{\Lambda }}_{i,i}^2, i\in{\cal L}$ and ${\bf{\bar \Lambda }} = {\rm{diag}}\left( {{{{\bf{\bar \Lambda }}}_{1,1}}, \ldots ,{{{\bf{\bar \Lambda }}}_{L,L}}} \right)$.  For a given  transmit beamforming vector $\bf w$, 
 the subproblem regarding pilot matrix  ${{{\bf{\bar \Lambda }}}}$ is given by

\begin{subequations} \label{subproblem2}
	\begin{align}
	& \mathop {\max }\limits_{{{{\bf{\bar \Lambda }}}_{i,i}} \ge 0} {N_r}\sum\limits_{i = 1}^L {{{\log }_2}\left( {1 + \frac{{\delta _g^2{{{\bf{\bar \Lambda }}}_{i,i}}}}{{\sigma _r^2}}} \right)}  + {N_r}\left( {{T_c} - L} \right){\log _2}\left( {1 + \frac{P_t{\delta _g^2}}{{\sigma _r^2}}} \right)\label{subproblem2_obj}\\
	& {\rm s.t.}~ {{\bf{w}}^H}{{\bf{R}}_{{\bf{ h}}}}{\bf{w}} \ge {2^{\frac{{{R_{{\rm{th}}}}{T_c}}}{{{T_c} - L}}}}\left( {\sigma _p^2{{\bf{w}}^H}{{\bf{U}}_{\bf{h}}}{\bf{\Sigma }}_{\bf{h}}^{\frac{1}{2}}{{\left( {{\bf{\Sigma }}_{\bf{h}}^{\frac{1}{2}}{{\bf{R}}_{\bf{\Lambda }}}{\bf{\Sigma }}_{\bf{h}}^{\frac{1}{2}} + \sigma _p^2{{\bf{I}}_{{N_t}}}} \right)}^{ - 1}}} \right.\notag\\
	&\qquad \times{{\bf{\Sigma }}_{\bf{h}}^{\frac{1}{2}}{\bf{U}}_{\bf{h}}^H{\bf{w}} + \sigma _t^2} \bigg) - \sigma _t^2,\label{subproblem2_const1} \\
	& \qquad \sum\limits_{i = 1}^L {{{{\bf{\bar \Lambda }}}_{i,i}}}  + \left( {{T_c} - L} \right){\left\| {\bf{w}} \right\|^2} \le {T_c}{P_{\rm  ave }}, \label{subproblem2_const2}
	\end{align}
\end{subequations}
where ${{\bf{R}}_{{\bf{\bar \Lambda }}}} = \left[ {\begin{array}{*{20}{c}}
	{{\bf{\bar \Lambda }}}&{{{\bf{0}}_{L \times \left( {{N_t} - L} \right)}}}\\
	{{{\bf{0}}_{\left( {{N_t} - L} \right) \times L}}}&{{{\bf{0}}_{{N_t} - L}}}
	\end{array}} \right]$.

To tackle  the inversion matrix involved in constraint \eqref{subproblem2_const1}, we first introduce an auxiliary variable $\eta \ge0 $  and  equivalently transform it into 
\begin{align}
&{{\bf{w}}^H}{{\bf{R}}_{\bf{h}}}{\bf{w}} \ge {2^{\frac{{{R_{{\rm{th}}}}{T_c}}}{{{T_c} - L}}}}\left( {\sigma _p^2\eta  + \sigma _t^2} \right) - \sigma _t^2, \label{P0_nonunitary_new_sub1_const1_v1_sub1}
\end{align}
and
\begin{align}
&\eta  \ge {{\bf{w}}^H}{{\bf{U}}_{\bf{h}}}{\bf{\Sigma }}_{\bf{h}}^{\frac{1}{2}}{\left( {{\bf{\Sigma }}_{\bf{h}}^{\frac{1}{2}}{{\bf{R}}_{\bf{\bar \Lambda }}}{\bf{\Sigma }}_{\bf{h}}^{\frac{1}{2}} + \sigma _p^2{{\bf{I}}_{{N_t}}}} \right)^{ - 1}}{\bf{\Sigma }}_{\bf{h}}^{\frac{1}{2}}{\bf{U}}_{\bf{h}}^H{\bf{w}}. \label{P0_nonunitary_new_sub1_const1_v1_sub2}
\end{align}
Then, by applying the Schur complement technique, we rewrite \eqref{P0_nonunitary_new_sub1_const1_v1_sub2} into a linear matrix inequality form given by 
\begin{align}
\left[ {\begin{array}{*{20}{c}}
	\eta &{{{{\bf{ w}}}^H}{{\bf{U}}_{\bf{h}}}{\bf{\Sigma }}_{\bf{h}}^{\frac{1}{2}}}\\
	{{\bf{\Sigma }}_{\bf{h}}^{\frac{1}{2}}{\bf{U}}_{\bf{h}}^H{\bf{ w}}}&{{\bf{\Sigma }}_{\bf{h}}^{\frac{1}{2}}{{{\bf{ R}}}_{\bf{\bar \Lambda }}}{\bf{\Sigma }}_{\bf{h}}^{\frac{1}{2}} + \sigma _p^2{{\bf{I}}_{{N_t}}}}
	\end{array}} \right] \succeq {{\bf{0}}_{{N_t} + 1}}.\label{P0_nonunitary_new_sub1_const1_v1_sub2_v1}
\end{align}
Based on \eqref{P0_nonunitary_new_sub1_const1_v1_sub1}, \eqref{P0_nonunitary_new_sub1_const1_v1_sub2}, and \eqref{P0_nonunitary_new_sub1_const1_v1_sub2_v1}, problem \eqref{subproblem2} can be recast as 
\begin{subequations} \label{subproblem2_new}
	\begin{align}
	& \mathop {\max }\limits_{\eta\ge 0,{{{\bf{\bar \Lambda }}}_{i,i}} \ge 0} {N_r}\sum\limits_{i = 1}^L {{{\log }_2}\left( {1 + \frac{{\delta _g^2{{{\bf{\bar \Lambda }}}_{i,i}}}}{{\sigma _r^2}}} \right)}  \notag\\
	&\qquad\qquad + {N_r}\left( {{T_c} - L} \right){\log _2}\left( {1 + \frac{P_t{\delta _g^2}}{{\sigma _r^2}}} \right)\label{subproblem2_new_obj}\\
	& {\rm s.t.}~ \eqref{subproblem2_const2}, \eqref{P0_nonunitary_new_sub1_const1_v1_sub1},   \eqref{P0_nonunitary_new_sub1_const1_v1_sub2_v1}.
	\end{align}
\end{subequations}
It can be readily verified that the objective function  is concave and all constraints are convex, and thus problem \eqref{subproblem2_new} can be solved by convex optimization solvers.

\begin{algorithm}[!t]
	\caption{The BCD algorithm  for solving  problem   \eqref{P0_nonunitary}.}
	\label{alg1}
	\begin{algorithmic}[1]
		\STATE  \textbf{Search} $L$ from $1$  to  $T_c$.
		\STATE  \quad \textbf{Initialize} ${{{{\bf{\bar \Lambda }}}_{i,i}}}, i \in {\cal L}$  and ${\varepsilon}$.
		\STATE  \quad \textbf{Repeat}
		\STATE  \qquad Update transmit beamforming vector ${\bf{w}}$ by solving\\
		\qquad  problem \eqref{subproblem1_new}.
		\STATE  \qquad Update pilot matrix ${{\bf{\bar \Lambda }}}$ by solving problem \eqref{subproblem2_new}.
		\STATE  \quad \textbf{End} the fractional increase of the objective value is less \\
		\quad  than ${\varepsilon}$.
		\STATE \textbf{End}
		\STATE  Pick up the  best solution that maximizes the objective function  among $T_c$  result solutions.
	\end{algorithmic}
\end{algorithm}
\subsubsection{Overall Algorithm and Computational Complexity}
Based on the solutions to the above subproblems, a BCD algorithm is proposed by
optimizing the two subproblems in an iterative way, where
the solution obtained in each iteration is used as the initial point
of the next iteration, which is summarized in Algorithm~\ref{alg1}.
Since at each iteration, each subproblem is optimally solved,  the obtained solution converges to a stationary point \cite{boyd2004convex}. The computational complexity of Algorithm~\ref{alg1} is $L{I_{{\rm{iter}}}}\left( {{{\left( {{N_t} + 1} \right)}^{3.5}} + {L^{3.5}}} \right)$ \cite{gondzio1996computational}, where ${I_{{\rm{iter}}}}$ denotes the number of iterations required to reach
convergence in the inner layer.

\section{Numerical Results}
In this section, we provide numerical results    to validate   the effectiveness of the proposed designs in the  ISAC system.  The large-scale path loss for the communication channel between the BS and the user is modeled as   ${L_{{\rm{loss}}}} = {L_0}{\left( {d/{d_0}} \right)^{ - \alpha }}$, where ${ L}_0$ denotes the channel power gain at the reference distance of  ${{d_0}}=1~\rm m$,  $d$ is the link distance, and $\alpha$ is the path loss exponent. The small-scale fading of the   BS-user link  follows an   exponentially correlated Rician  channel   \cite{zhao2021intelligent}
\begin{align}
{\bf{h}} = \sqrt {\frac{{{\beta _{{\rm{rician}}}}}}{{1 + {\beta _{{\rm{rician}}}}}}} {{\bf{h}}^{{\rm{LoS}}}} + \sqrt {\frac{1}{{1 + {\beta _{{\rm{rician}}}}}}} {{\bf{\Phi }}^{\frac{1}{2}}}{{\bf{h}}^{{\rm{NLoS}}}},
\end{align}
where ${{\beta _{{\rm{rician}}}}}$ stands for the Rician factor, ${\bf  \Phi}$ denotes the  spatial correlation matrix between the BS and the user, ${{\bf{h}}^{{\rm{LoS}}}}$ denotes the deterministic line-of-sight (LoS),  and  the entries of ${{\bf{h}}^{{\rm{NLoS}}}}$  are assumed to be independent and identically distributed and follow a CSCG distribution with zero mean and unit variance. 
We consider the
following exponential correlation model of ${\bf{\Phi }}$ \cite{loyka2001channel}
\begin{align}
{{\bf  \Phi}}\left( {i,j} \right) = \left\{ \begin{array}{l}
r^{j - i},{\rm{if}}{\kern 1pt} {\kern 1pt} {\kern 1pt} {\kern 1pt} {\kern 1pt} i \le j,\\
{{\bf  \Phi}}\left( {j,i} \right),{\rm{if}}{\kern 1pt} {\kern 1pt} {\kern 1pt} {\kern 1pt} {\kern 1pt} i > j,
\end{array} \right.
\end{align}
where $0 \le {r} \le 1$ denotes  the correlation coefficient. A large $r$ means that  the channel is more spatially correlated and $r=0$ indicates that the channel is  uncorrelated. The second-order statistic of ${\bf{h}}$ is given by ${{\bf{R}}_{\bf{h}}} = {\mathbb E}\left\{ {{\bf{h}}{{\bf{h}}^H}} \right\} = \frac{{{\beta _{{\rm{rician}}}}}}{{1 + {\beta _{{\rm{rician}}}}}}{{\bf{h}}^{{\rm{LoS}}}}{{\bf{h}}^{{\rm{LoS}},{\rm{H}}}} + \frac{1}{{1 + {\beta _{{\rm{rician}}}}}}{\bf{\Phi }}$. 
Without loss of generality, we assume that all the noise powers  are the same, i.e., $\sigma_t ^2=\sigma_p ^2=\sigma_r ^2=\sigma ^2$.
 Unless  specified otherwise, we set   $N_t=20$, $N_r=4$,   ${P_{{\rm{ave}}}} = 30~{\rm dBm}$, $T_c=15$, ${P_{{\rm{fa}}}} = {10^{ - 6}}$, $r=0.6$, $\beta_{\rm rician}=0~{\rm dB}$, $\delta _g^2 =  - 100~{\rm dBm}$, $d=200~{\rm m}$, $\alpha=3.2$, $\sigma^2=-80~{\rm dBm}$, and $\varepsilon  = {10^{ - 4}}$.


\begin{figure*}[!t]
	\centering
	\subfigure[Unlimited power case: $P_{\rm tot}=LP_{\rm ave}$.]{
		\begin{minipage}[t]{0.45\linewidth}
			\centering
			\includegraphics[width=2.8in]{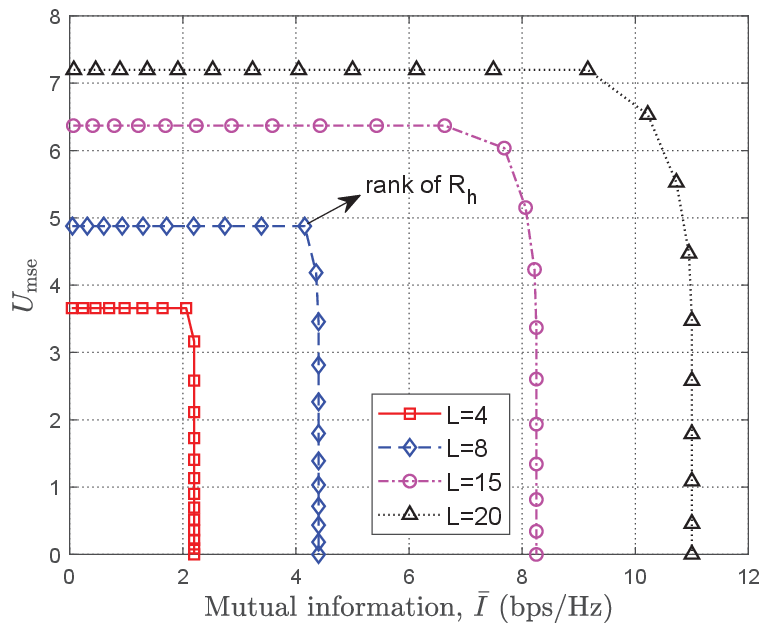}
		\end{minipage}%
	}%
	\hspace{25pt}           
	\subfigure[Limited power case: $P_{\rm tot}=T_cP_{\rm ave}$.]{
		\begin{minipage}[t]{0.45\linewidth}
			\centering
			\includegraphics[width=2.7in]{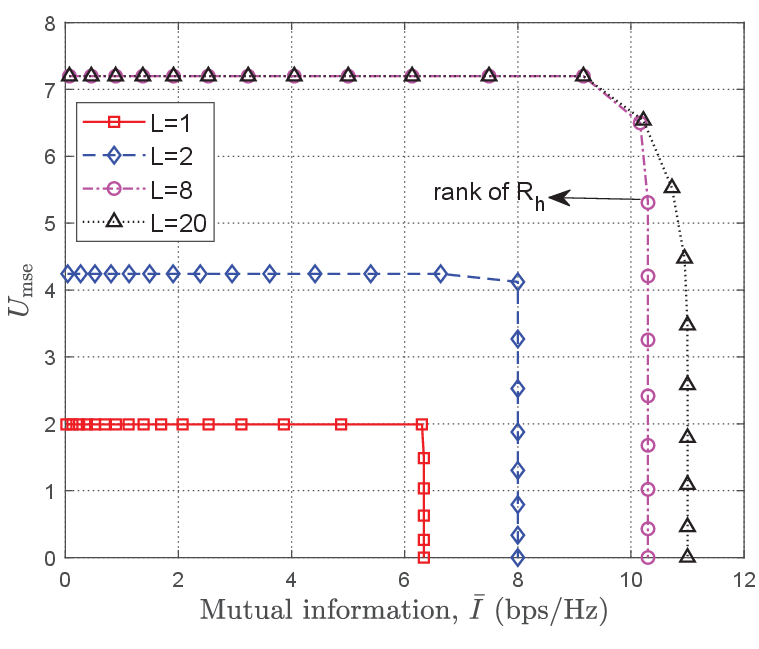}
		\end{minipage}
	}%
	\quad
	\centering
	\caption{MSE-MI region for the different length of  pilot sequences  $L$ under $\delta _g^2 =  - 90~{\rm dBm}$.}\label{MSE-MI}
\end{figure*}

\subsection{MSE-MI Region of ISAC System}
To evaluate the MSE-MI region, we  introduce an  utility function to describe the normalized   MSE of  channel $\bf h$
denoted by ${U_{{\rm{mse}}}} = \frac{1}{{{h_{{\rm{NMSE}}}}}}$, where ${{h_{{\rm{NMSE}}}}}$ is given by 
\begin{align}
{h_{{\rm{NMSE}}}} &= \frac{{{\mathbb E}\left\{ {{{\left\| {{\bf{h}} - {\bf{\hat h}}} \right\|}^2}} \right\}}}{{{\mathbb E}\left\{ {{{\left\| {\bf{h}} \right\|}^2}} \right\}}} \notag\\
&= \frac{{\sigma _p^2{\rm{tr}}\left( {{{\bf{R}}_{\bf{h}}}{{\left( {{\bf{X}}_c^H{{\bf{X}}_c}{{\bf{R}}_{\bf{h}}} + \sigma _p^2{{\bf{I}}_{{N_t}}}} \right)}^{ - 1}}} \right)}}{{{\rm{tr}}\left( {{{\bf{R}}_{\bf{h}}}} \right)}}. \label{normized_channel}
\end{align}
It can be seen  that a smaller  ${{h_{{\rm{NMSE}}}}}$ leads to   a larger utility function value ${U_{{\rm{mse}}}}$.

Then, the  MSE-MI region is defined as 
\begin{align}
&{C_{\bar I - {{{U}}_{{\rm{mse}}}}}} = \left\{ {\left( {\bar I,{{{U}}_{{\rm{mse}}}}} \right):\bar I \le {N_r}{{\log }_2}\left| {{{\bf{I}}_L}{\rm{ + }}\frac{{\delta _g^2}}{{\sigma _r^2}}{{\bf{X}}_c}{\bf{X}}_c^H} \right|,} \right.\notag\\
&\!\!\left. {{U_{{\rm{mse}}}} \le \frac{{{\rm{tr}}\left( {{{\bf{R}}_{\bf{h}}}} \right)}}{{\sigma _p^2{\rm{tr}}\left( {{{\bf{R}}_{\bf{h}}}{{\left( {{\bf{X}}_c^H{{\bf{X}}_c}{{\bf{R}}_{\bf{h}}} + \sigma _p^2{{\bf{I}}_{{N_t}}}} \right)}^{ - 1}}} \right)}},\left\| {{{\bf{X}}_c}} \right\|_F^2 \le {P_{{\rm{tot}}}}} \right\},
\end{align}
where ${{P_{{\rm{tot}}}}}$ stands for the power budget.  The boundary of ${C_{\bar I - {{{U}}_{{\rm{mse}}}}}}$ is called the Pareto boundary, which consists
of all the ${\bar I}$\text{-}${U_{{\rm{mse}}}}$ tuples at which it is impossible to increase ${U_{{\rm{mse}}}}$
without simultaneously decreasing  MI, and vice versa \cite{jorwieck2008Complete}.

In Fig.~\ref{MSE-MI} (a), we show the  MSE-MI region with unlimited power constraint, i.e., ${P_{{\rm{tot}}}} = L{P_{{\rm{ave}}}}$, which indicates that the available power budget is monotonically increasing with $L$. It is observed that the MSE-MI region enlarges as $L$  increases. This is expected since more power can be used for both channel estimation and  target detection as $L$ increases. In addition, 
it is observed that  the  extreme points  on the vertical axis, i.e., the points where the curves  intersect the vertical axis,  increase with $L$, which  is consistent with Remark $1$. Meanwhile, we can see that the  extreme points  on the horizontal  axis also increase with $L$, which is consistent with Remark $2$.  Moreover, we can clearly see that there exists a tradeoff between  maximizing MI  and maximizing ${U_{{\rm{mse}}}}$.
Furthermore, we show in Fig.~\ref{MSE-MI} (b) that the  MSE-MI region with the limited power constraint, i.e., ${P_{{\rm{tot}}}} = T_c{P_{{\rm{ave}}}}$, where ${P_{{\rm{tot}}}}$ is fixed and independent of $L$. It can be observed that the MSE-MI region enlarges when $L \le {\rm{rank}}\left( {{{\bf{R}}_{\bf{h}}}} \right)$, while the ${U_{{\rm{mse}}}}$ will not increase as $L \ge {\rm{rank}}\left( {{{\bf{R}}_{\bf{h}}}} \right)$ (see the curves corresponding to $L=8$ and $L=20$). This is because the MSE of the channel will not be further reduced as  $L \ge {\rm{rank}}\left( {{{\bf{R}}_{\bf{h}}}} \right)$ when the limited power constraint is considered, which is explicitly unveiled in Remark 1.


\begin{figure}[!t]
	\centerline{\includegraphics[width=3.2in]{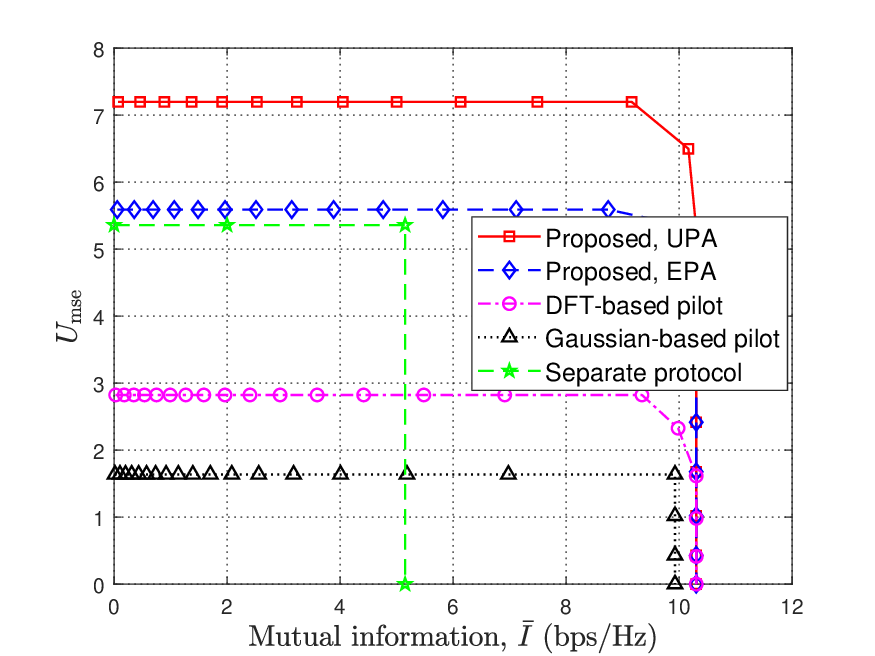}}
	\caption{MSE-MI region comparison for different pilot design approaches  under $L=8$ and $\delta _g^2 =  - 90~{\rm dBm}$.} \label{MSE-MI-pilots}
	\vspace{-0.5cm}
\end{figure}

\begin{figure}[!t]
	\centerline{\includegraphics[width=3.2in]{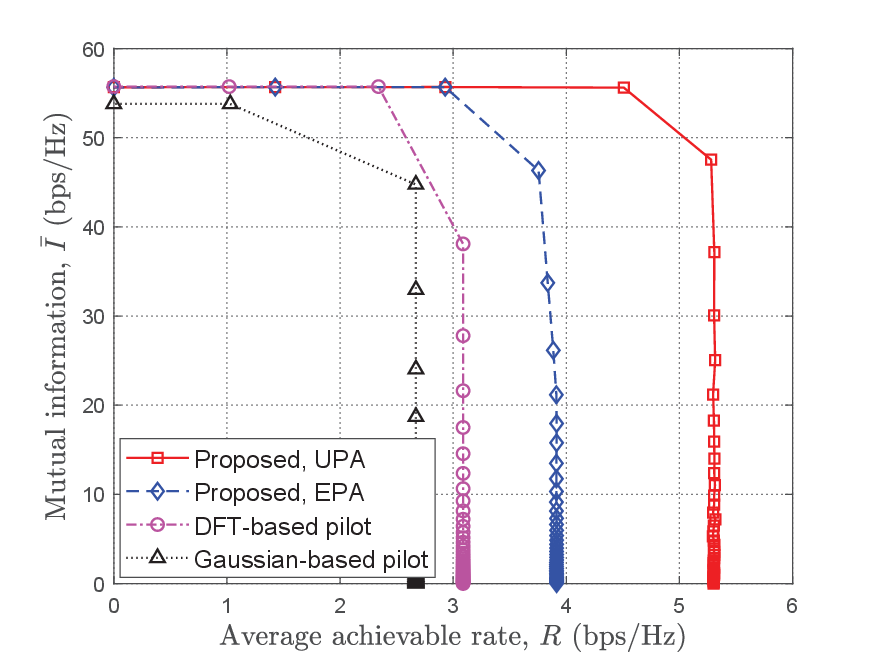}}
	\caption{Rate-MI region comparison for different pilot design approaches  under   $\delta _g^2 =  - 100~{\rm dBm}$ and ${P_{{\rm{ave}}}} = 40~{\rm dBm}$.} \label{RvsMI_differentpilots}
	\vspace{-0.5cm}
\end{figure}

To show the superiority of the proposed pilot structure, we consider the following approaches for comparison.
\begin{itemize}
	\item \textbf{Proposed, UPA (unequal power allocation):} This is our proposed pilot structure, i.e., ${\bf{X}}_c^H = {{\bf{U}}_{\bf{h}}}\left( {1:L} \right){\bf{\Lambda }}$, where ${\bf{\Lambda }} = {\rm{diag}}\left( {{{\bf{\Lambda }}_{1,1}}, \ldots ,{{\bf{\Lambda }}_{L,L}}} \right)$ with ${{\bf{\Lambda }}_{i,i}}\ge 0$ needs to be optimized. 
	
	\item \textbf{Proposed, EPA (equal power allocation): } It is the same as the above scheme except that  the power allocation  in the main diagonal of ${\bf{\Lambda }}$ is  equal, i.e., ${{\bf{\Lambda }}_{1,1}} =  \ldots  = {{\bf{\Lambda }}_{L,L}}$, such that ${{\bf{X}}_c}{\bf{X}}_c^H$ is orthogonal. 
	\item \textbf{DFT (discrete Fourier transform)-based pilot:} The entries in ${{\bf{X}}_c}$ are given by ${\left[ {{\bf{X}}_c} \right]_{i,j}} = \sqrt{P_{\rm DFT}}{e^{ - j\frac{{2\pi \left( {i - 1} \right)\left( {j - 1} \right)}}{{{N_t}}}}},1 \le i \le L,1 \le j \le {N_t}$, where $P_{\rm DFT}$ represents the allocated power. We have   ${{\bf{X}}_c}{\bf{X}}_c^H = {P_{{\rm{DFT}}}}{{\bf{I}}_L}$, which is orthogonal. 
	\item \textbf{Gaussian-based pilot:}  Each entry in ${{\bf{X}}_c}$ is independent of each other  and follows ${\left[ {{\bf{X}}_c} \right]_{i,j}} \sim {\cal CN}\left( {0,{P_{{\rm{Guassian}}}}} \right),1 \le i \le L,1 \le j \le {N_t}$, where ${{P_{{\rm{Guassian}}}}}$ represents the  power.
	\item \textbf{Separate protocol:}  Stage I is equally divided into two sub-stages. One is for target  detection and the other is for channel estimation. The optimal  matrices for maximizing the MI and $U_{\rm mse}$ can be directly  obtained based on  Section III, respectively.
\end{itemize}

In Fig.~\ref{MSE-MI-pilots}, we study the MSE-MI tradeoff for different pilot design approaches. It is observed that the
MSE-MI region obtained by our proposed pilot structure with UPA is significantly larger than those of the other approaches, which demonstrates the superiority of the proposed pilot design. In addition, we observe that the
MSE-MI region obtained by our proposed pilot structure with UPA is larger than that obtained by our proposed pilot structure with EPA since  allocating unequal power on pilot sequences  can further reduce  channel estimation error (equivalently increase ${U_{{\rm{mse}}}}$) shown in \eqref{optimal_solution_Lba}. Compare 
the ``Proposed, EPA'' approach  to the  DFT-based pilot approach, both pilot structures are orthogonal, whereas the MSE-MI region obtained by the  ``Proposed, EPA'' approach is larger than that obtained by the DFT-based pilot approach. The reason is that  the  ``Proposed, EPA'' approach exploits the  communication channel information for channel estimation so that the channel estimation error can be further reduced.  Furthermore, the Gaussian-based pilot achieves the smallest MSE-MI region since  the Gaussian-based pilot is neither orthogonal  nor exploiting the communication channel information.  It is worth pointing out that    all the approaches except the Gaussian-based pilot approach achieves the same maximum MI  since the orthogonal pilot matrix is optimal for maximizing the MI.   Moreover,  compared to   ``Separate protocol'', we can observe that our proposed  scheme significantly outperforms it.
This is  due to the following two reasons. First, the full time duration can be utilized for the proposed scheme, while only a part of the time duration can be utilized either for target detection or channel estimation. Second, the proposed unified pilot matrix can balance the target detection and channel estimation.

\begin{figure}[!t]
	\centerline{\includegraphics[width=3.2in]{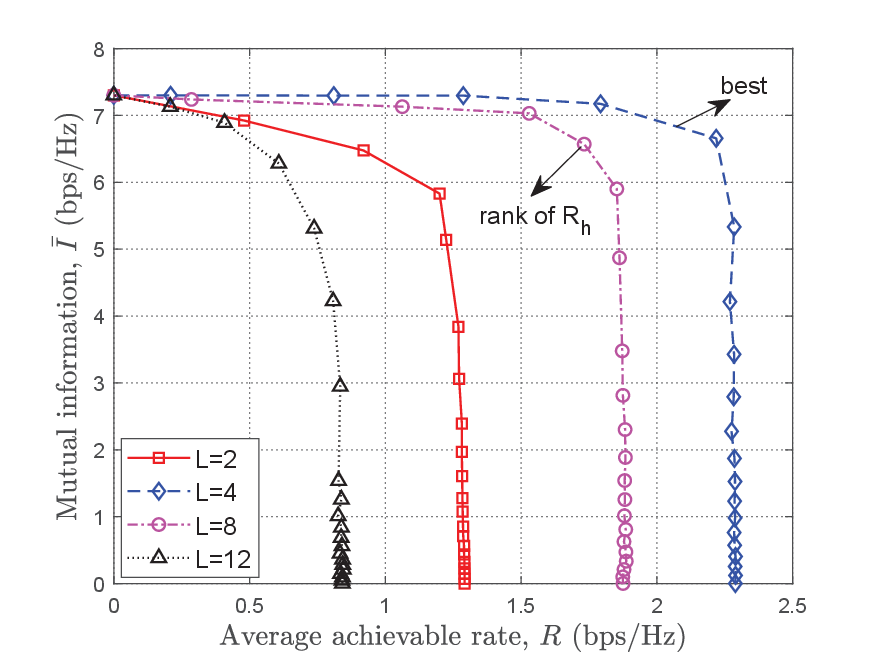}}
	\caption{Rate-MI region comparison for the different length of  pilot sequences  $L$ under $\delta _g^2 =  - 90~{\rm dBm}$.} \label{RvsMI}
	\vspace{-0.5cm}
\end{figure}

\subsection{Achievable Rate-MI Region of ISAC System}
Before proceeding to describe the Rate-MI  region, we first define the feasible region of optimization variables $\left\{ {{\bf{w}},L,{{\bf{X}}_c}} \right\}$ as 
\begin{align}
\!\!\!\Omega   = \left\{ {\frac{{\left\| {{{\bf{X}}_c}} \right\|_F^2 + \left( {{T_c} - L} \right){{\left\| {\bf{w}} \right\|}^2}}}{{{T_c}}} \le {P_{{\rm{ave}}}},0 \le L \le {T_c},L \in {{\mathbb Z}^ + }} \right\}.
\end{align}
As a result, the achievable Rate-MI  region is defined as 
\begin{align}
&\!\!\!{{\cal C}_{R - \bar I}} \overset{\triangle}= \left\{ {\left( {R,\bar I} \right):\bar I \le {N_r}{{\log }_2}\left| {{{\bf{I}}_L}{\rm{ + }}\frac{{\delta _g^2}}{{\sigma _r^2}}{{\bf{X}}_c}{\bf{X}}_c^H} \right|} \right. \notag\\
&\qquad \qquad\qquad\quad+ {N_r}\left( {{T_c} - L} \right){\log _2}\left( {1 + \frac{{\delta _g^2{{\left\| {\bf{w}} \right\|}^2}}}{{\sigma _r^2}}} \right),\notag\\
&\!\!\!\!\left. {R \le \frac{{{T_c} - L}}{{{T_c}}}{{\log }_2}\left( {\frac{{{{\bf{w}}^H}{{\bf{R}}_{\bf{h}}}{\bf{w}} + \sigma _t^2}}{{{{\bf{w}}^H}{{\bf{R}}_{{{\bf{h}}_e}}}{\bf{w}} + \sigma _t^2}}} \right),{\bf{w}} \in \Omega ,L \in \Omega ,{{\bf{X}}_c} \in \Omega } \right\},
\end{align}
where  the $R$-$\bar I$ tuples  on the Pareto boundary of ${\cal C}_{R-{\bar I}}$ can be similarly obtained by Algorithm~\ref{alg1}.

In Fig.~\ref{RvsMI_differentpilots}, we compare the  Rate-MI tradeoff for different pilot approaches. It is observed that 
the Rate-MI region obtained by  our proposed pilot structure with UPA is larger than the other  approaches, which demonstrates the  benefit of joint design of the transmit beamformer and pilot sequences. In addition, we observe that the MI is the same for all approaches except the Gaussian-based pilot approach when the required achievable rate is small. This is because   as $R$ is small, the minimum rate required by the  user can be readily satisfied  by just allocating equal power  in the pilot training stage and information transmission stage so that the obtained  MI $\bar I$ is the same.
However, for the Gaussian-based pilot approach, since this pilot structure is not orthogonal, the performance of MI will be impaired and thus,  a  performance degradation is incurred.  


\begin{figure}[!t]
	\centerline{\includegraphics[width=3.2in]{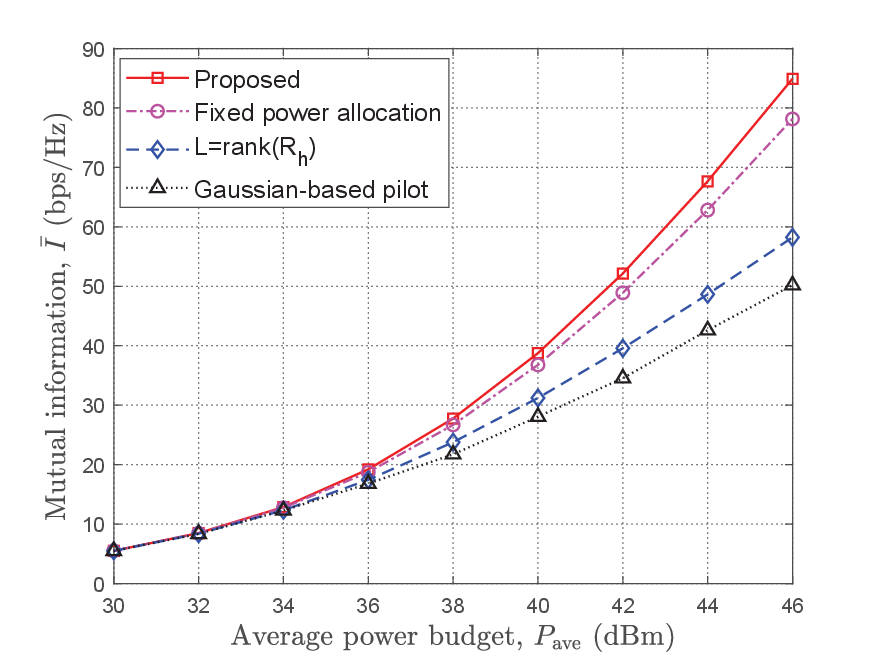}}
	\caption{$\bar I$ versus $P_{\rm ave}$ under $R_{\rm th}=2 ~{\rm bps/Hz}$, $\beta_{\rm rician}  =  - \infty $ (in dB), $d=100~{\rm m}$,  $r=0.8$, and  $\alpha=2.6$.} \label{MIvsP_max}
	\vspace{-0.5cm}
\end{figure}

In Fig.~\ref{RvsMI}, we compare the Rate-MI tradeoff for different $L$. It is observed that the Rate-MI region firstly enlarges  as $L$ increases (see curves from $L=2$ to $L=4$), and then shrinks when $L$ becomes large (see curves from $L=8$ to $L=12$). This is because  the effective  spectral efficiency of the user is affected by two factors: 1) the  duration time $L$ for channel estimation and 2) the  duration time $T_c-L$  for information transmission. As $L$ is small,  the channel estimation error is large, and the achievable rate is  small even if the remaining duration time, i.e., $T_c-L$, for information transmission is large.  In contrast, as $L$ becomes large,  the channel estimation error is significantly reduced, whereas the achievable rate is still small since the remaining time for information transmission is small.  Interestingly, we can see that $L{\rm{ = rank}}\left( {{{\bf{R}}_{\bf{h}}}} \right)$ is not the optimal pilot sequence length  to maximize the Rate-MI region. 
Therefore, there exists a tradeoff between minimizing the channel estimation error and maximizing the user's effective throughput.

\begin{figure}[!t]
	\centerline{\includegraphics[width=3.2in]{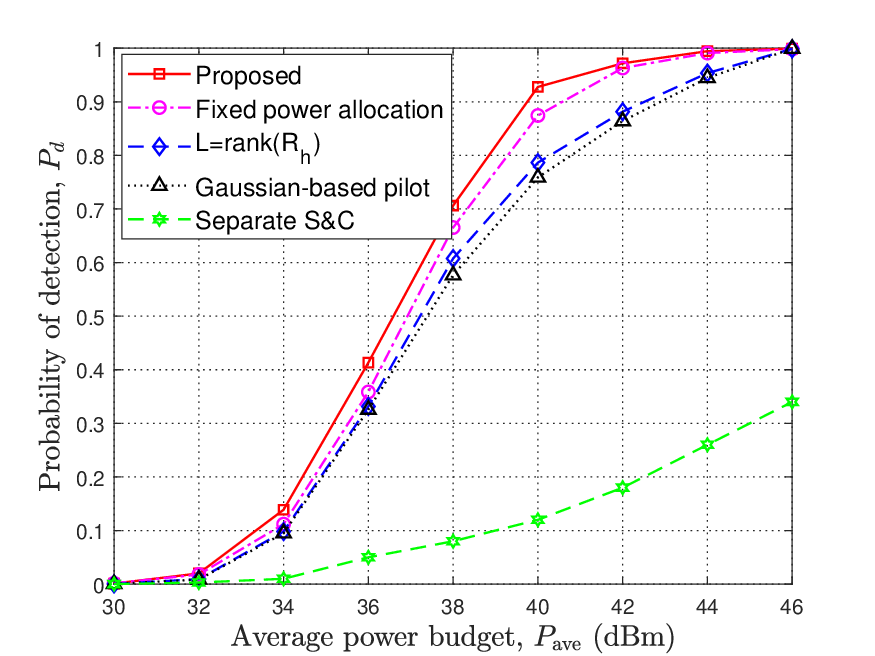}}
	\caption{$P_d$ versus $P_{\rm ave}$ under $R_{\rm th}=2 ~{\rm bps/Hz}$, $\beta_{\rm rician}  =  - \infty $ (in dB), $d=100~{\rm m}$,  $r=0.8$, and  $\alpha=2.6$.} \label{P_dvsP_max}
	\vspace{-0.5cm}
\end{figure}

In Fig.~\ref{MIvsP_max}, we compare the MI obtained by different approaches versus $P_{\rm ave}$. The following benchmark approaches are compared: 1) \textbf{Fixed power allocation:} Similar to the proposed approach except that  the power allocated to two stages, i.e., the channel estimation stage and the information transmission stage, is the same; 2) $L{\rm{ = rank}}\left( {{{\bf{R}}_{\bf{h}}}} \right)$: Similar to the proposed approach except that  the pilot length is fixed with $L{\rm{ = rank}}\left( {{{\bf{R}}_{\bf{h}}}} \right)$; 3) \textbf{Gaussian-based pilot:} Similar to the proposed approach except that  each entry in   ${\bf{X}}_c^H$ is CSCG distributed. 4)\textbf{ Separate S\&C:} 
The coherence time  is divided into three stages, namely target detection, channel estimation, and information transmission.  The optimal sensing matrix for target detection and the optimal pilot  matrix  for channel estimation can be easily derived in closed-form expressions, while the beamformer can be similarly solved by the proposed Algorithm 1.
It is observed that the MI for all approaches increases monotonically with $P_{\rm ave}$ as expected. In addition, we observe that the proposed approach outperforms the other benchmark approaches, which indicates the benefits of joint design of pilot design and transmit beamformer. 
To show the impact of MI on the target detection performance, the corresponding $P_d$ versus $P_{\rm ave}$ is studied in  Fig.~\ref{P_dvsP_max}.  We first obtain the solution based on the MI maximization optimization problem, and then  substitute  the obtained  pilot  matrix, transmit beamformer, and training duration into \eqref{P_obj}  to obtain the corresponding target detection probability.  Compared to Fig.~\ref{MIvsP_max}, we observe that a higher MI corresponds to  a higher target  detection probability, which demonstrates that maximizing the MI potentially increases the  detection probability. Moreover, we  can observe that the ``Separate S\&C" scheme achieves very low detection probability even as  power budget is large. This is because only a part of the time is utilized for target detection. In contrast,
	our proposed scheme achieves much higher detection probability than the ``Separate S\&C" scheme, which implies that the proposed  protocol outperforms the separate  protocol. The reason is that full-time duration  and a well-customized pilot matrix are leveraged in our proposed scheme.


\begin{figure}[!t]
	\centerline{\includegraphics[width=3.2in]{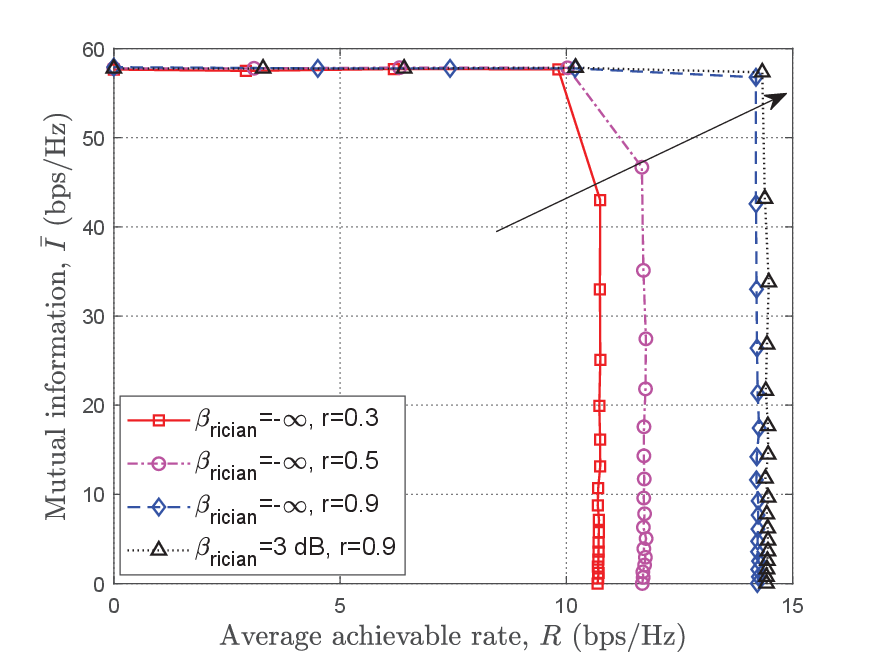}}
	\caption{Rate-MI region for different correlation coefficients  $\left\{ {r = 0.3,r = 0.5,r = 0.9} \right\}$  under  ${P_{{\rm{ave}}}} = 40~{\rm dBm}$,  $\sigma^2=-90~{\rm dBm}$,  $d=100~{\rm m}$, and a path loss exponent of $2.6$.} \label{channel_correlation}
	\vspace{-0.5cm}
\end{figure}

In Fig.~\ref{channel_correlation}, we study the Rate-MI tradeoff for different correlation coefficients $r$ under $\beta_{\rm rician}  =  - \infty $ (in dB) and $\beta_{\rm rician}  =  3~{\rm dB} $. It is observed that the Rate-MI region enlarges as $r$ increases. This is because   as $r$ increases, the communication channel becomes more spatially correlated and the rank of its covariance matrix, i.e., ${\rm{rank}}\left( {{{\bf{R}}_{\bf{h}}}} \right)$, becomes smaller. Therefore, the need of the length of pilot sequences for channel estimation is reduced and the remaining duration for information transmission increases, and the effective spectral efficiency of user is thus increased. In addition, with the same correlation coefficient, i.e., $r=0.9$, a larger ${{\beta _{{\rm{rician}}}}}$ leads to a larger Rate-MI region. This is because as ${{\beta _{{\rm{rician}}}}}$ increases,  the communication channel becomes more spatially correlated and the rank of ${\rm{rank}}\left( {{{\bf{R}}_{\bf{h}}}} \right)$ is also reduced so that the effective spectral efficiency of user improves.

\section{Conclusion}
In this paper, we investigated the performance limit of ISAC by studying the MSE-MI and Rate-MI regions. We proposed a  novel target detection and information transmission protocol where  both channel estimation and information transmission stages are leveraged for target detection. The corresponding  target detection probability  was derived by applying the GLRT-based detector. By respectively designing the optimal pilot matrix for the  channel estimation and target detection, the fundamental tradeoff between minimizing the channel estimation error and maximizing  MI was unveiled and a novel pilot structure was then proposed to balance the above two functionalities. In addition, the impact of training duration on  channel estimation and target detection was characterized.
Finally, we proposed an efficient iterative algorithm to maximize MI by jointly optimizing the  pilot matrix, the training duration, and the transmit beamformer.
Extensive simulation results under various practical setups demonstrated that our proposed pilot structure can well balance the system performance between the communication transmission  and the  target detection, and by jointly optimizing the pilot matrix and transmit beamformer, the Rate-MI region can be significantly enlarged.
Furthermore, it was also unveiled that as the  communication channel is  more spatially correlated, 
 the Rate-MI region can be further enlarged.

\appendices
\section{Proof of Lemma~2} \label{appendix1}
Let ${\bf{X}} = {{\bf{U}}_{\bf{x}}}{{\bf{\Sigma }}_{\bf{x}}}{\bf{V}}_{\bf{x}}^H$ by performing (reduced) singular value decomposition on ${\bf{X}}$, where ${{\bf{U}}_{\bf{x}}} \in {{\mathbb  C}^{{T_c} \times \upsilon }},{{\bf{\Sigma }}_{\bf{x}}} \in {\mathbb C^{\upsilon  \times \upsilon }},{\bf{V}}_{\bf{x}}^H \in {\mathbb C^{\upsilon  \times {N_t}}}$, and $\upsilon  = \min \left( {L + 1,{N_t}} \right)$,  with ${\bf{U}}_{\bf{x}}^H{{\bf{U}}_{\bf{x}}} = {{\bf{I}}_{{T_c}}}$, ${\bf{V}}_{\bf{x}}^H{{\bf{V}}_{\bf{x}}} = {{\bf{I}}_{{N_t}}}$, ${{\bf{\Sigma }}_{\bf{x}}} = {\rm{diag}}\left( {{{\bf{\Sigma }}_{1,1}}, \ldots ,{{\bf{\Sigma }}_\upsilon }} \right)$,  and ${{\bf{\Sigma }}_{1,1}} \ge  \ldots {{\bf{\Sigma }}_{\upsilon ,\upsilon }} > 0$. 
Thus, we have ${\left( {{{\bf{X}}^H}{\bf{X}}} \right)^\dag } = {{\bf{V}}_{\bf{x}}}{\bf{\Sigma }}_{\bf{x}}^{ - 2}{\bf{V}}_{\bf{x}}^H$. Then, substituting it into ${\rm{rank}}\left( {{\bf{X}}{{\left( {{{\bf{X}}^H}{\bf{X}}} \right)}^\dag }{{\bf{X}}^H}} \right)$ yields
\begin{align}
&{\rm{rank}}\left( {{\bf{X}}{{\left( {{{\bf{X}}^H}{\bf{X}}} \right)}^\dag }{{\bf{X}}^H}} \right)  \notag\\
&= {\rm{rank}}\left( {{{\bf{U}}_{\bf{x}}}{{\bf{\Sigma }}_{\bf{x}}}{\bf{V}}_{\bf{x}}^H{{\bf{V}}_{\bf{x}}}{\bf{\Sigma }}_{\bf{x}}^{ - 2}{\bf{V}}_{\bf{x}}^H{{\bf{V}}_{\bf{x}}}{{\bf{\Sigma }}_{\bf{x}}}{\bf{U}}_{\bf{x}}^H} \right)\notag\\
 &= {\rm{rank}}\left( {{{\bf{U}}_{\bf{x}}}{\bf{U}}_{\bf{x}}^H} \right)= {\rm{rank}}\left( {{{\bf{U}}_{\bf{x}}}} \right) = \min \left( {L + 1,{N_t}} \right) \notag\\
 &\overset{\triangle}{=} {\rm{rank}}\left( {\bf{X}} \right).
\end{align}
 This thus completes   the proof of Lemma~$2$.
 
\section{Proof of Lemma~3} \label{appendix2} 
We first check ${\bf{X}}{\left( {{{\bf{X}}^H}{\bf{X}}} \right)^\dag }{{\bf{X}}^H}$ is idempotent, i.e., ${\left( {{\bf{X}}{{\left( {{{\bf{X}}^H}{\bf{X}}} \right)}^\dag }{{\bf{X}}^H}} \right)^2} = {\bf{X}}{\left( {{{\bf{X}}^H}{\bf{X}}} \right)^\dag }{{\bf{X}}^H}$. Recall that ${\left( {{{\bf{X}}^H}{\bf{X}}} \right)^\dag } = {{\bf{V}}_{\bf{x}}}{\bf{\Sigma }}_{\bf{x}}^{ - 2}{\bf{V}}_{\bf{x}}^H$ (see it in Appendix A), we can obtain 
${\bf{X}}{\left( {{{\bf{X}}^H}{\bf{X}}} \right)^\dag }{{\bf{X}}^H} = {{\bf{U}}_{\bf{x}}}{\bf{U}}_{\bf{x}}^H$. Then,
we can express  ${\left( {{\bf{X}}{{\left( {{{\bf{X}}^H}{\bf{X}}} \right)}^\dag }{{\bf{X}}^H}} \right)^2}$ as 
\begin{align}
{\left( {{\bf{X}}{{\left( {{{\bf{X}}^H}{\bf{X}}} \right)}^\dag }{{\bf{X}}^H}} \right)^2}&= {{\bf{U}}_{\bf{x}}}{{\bf{\Sigma }}_{\bf{x}}}{\bf{V}}_{\bf{x}}^H{{\bf{V}}_{\bf{x}}}{\bf{\Sigma }}_{\bf{x}}^{ - 2}{\bf{V}}_{\bf{x}}^H{{\bf{V}}_{\bf{x}}}{{\bf{\Sigma }}_{\bf{x}}}{\bf{U}}_{\bf{x}}^H  \notag\\
&\times{{\bf{U}}_{\bf{x}}}{{\bf{\Sigma }}_{\bf{x}}}{\bf{V}}_{\bf{x}}^H{{\bf{V}}_{\bf{x}}}{\bf{\Sigma }}_{\bf{x}}^{ - 2}{\bf{V}}_{\bf{x}}^H{{\bf{V}}_{\bf{x}}}{{\bf{\Sigma }}_{\bf{x}}}{\bf{U}}_{\bf{x}}^H\notag\\
& = {{\bf{U}}_{\bf{x}}}{\bf{U}}_{\bf{x}}^H,
\end{align}
which indicates that ${\left( {{\bf{X}}{{\left( {{{\bf{X}}^H}{\bf{X}}} \right)}^\dag }{{\bf{X}}^H}} \right)^2} = {\bf{X}}{\left( {{{\bf{X}}^H}{\bf{X}}} \right)^\dag }{{\bf{X}}^H}$ and is thus  idempotent.

Next, we prove that the eigenvalue  of ${\bf{X}}{\left( {{{\bf{X}}^H}{\bf{X}}} \right)^\dag }{{\bf{X}}^H}$
is either  $0$ or $1$. Let ${\bf{M}} = {\bf{X}}{\left( {{{\bf{X}}^H}{\bf{X}}} \right)^\dag }{{\bf{X}}^H}$ and $\lambda$ and $\bf a$ (${\bf{a}} \ne {\bf{0}}$) be its eigenvalue and eigenvector, respectively. As such, we have 
\begin{align}
{\bf{Ma}} = \lambda {\bf{a}} \Rightarrow {{\bf{M}}^2}{\bf{a}} = \lambda {\bf{Ma}} \Rightarrow \lambda \left( {1 - \lambda } \right){\bf{a}} = {\bf{0}}.
\end{align}
Thus, $\lambda $ is equal to  either $0$ or $1$. In addition, since ${\rm{rank}}\left( {{\bf{X}}{{\left( {{{\bf{X}}^H}{\bf{X}}} \right)}^\dag }{{\bf{X}}^H}} \right) = {\rm{rank}}\left( {\bf{X}} \right) = \min \left( {L + 1,{N_t}} \right)$, the number of non-zeros is $\min \left( {L + 1,{N_t}} \right)$. Based on these, we  complete   the proof of Lemma~$3$.
\bibliographystyle{IEEEtran}
\bibliography{Estimation_transmission.bib}
\end{document}